| | |
|---|---|
| Title: | Drive Asymmetry, Convergence and the Origin of Turbulence in ICF Implosions |
| Author(s): | Vincent A. Thomas and Robert J. Kares<br>Applied Computational Physics (XCP) Division<br>Los Alamos National Laboratory |
| Published in: | *Coarse Grained Simulation and Turbulent Mixing*, edited by F. Grinstein, ISBN 978-1-107-13704-2, Cambridge University Press, 282 (2016). |

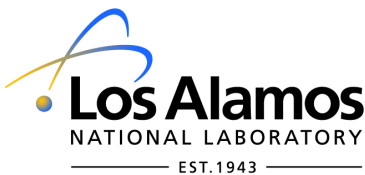

A note to the reader:
Figs. 3(a), 5(c), 13(a), 14(a), 14(b), 18(a) and 18(b) contain embedded hyperlinks to YouTube videos. Please point to
these figures and click to play the videos.

# Drive Asymmetry, Convergence and the Origin of Turbulence in ICF Implosions

*Vincent A. Thomas and Robert J. Kares*

*Los Alamos National Laboratory , Los Alamos, New Mexico*

**2D and 3D numerical simulations with the adaptive mesh refinement Eulerian radiation-hydrocode RAGE are used to investigate hydrodynamic disruption of asymmetrically driven ICF implosions. A central aspect of this phenomenon is the connection between drive asymmetry and the generation of turbulence in the DT fuel. Long wavelength deviations from spherical symmetry in the pressure drive lead to the generation of coherent vortical structures in the DT gas and it is the three dimensional instability of these structures that in turn leads to turbulence and mix. RAGE simulations with spatial resolutions as high as 0.05 μm in 3D are presented to exhibit the detailed mechanisms of turbulence growth. These simulations suggest that the amplification of small initial surface imperfections by acceleration-induced instabilities is not the only important source of turbulent mix in ICF implosions as is commonly supposed. Rather, the three dimensional instability of coherent vortical structures induced in the gas by the asymmetries in the implosion are an additional important source of turbulent mixing in ICF, perhaps even the dominant source. The effect of convergence on the hydrodynamic disruption of an asymmetrically driven ICF capsules is also considered, demonstrating how higher convergence is expected to lead to greater hydrodynamic disruption by both large scale fingers of pusher material into the fuel as well as the formation of radially outgoing turbulent jets of fuel. Implications of these results for NIF ignition are discussed.**

## 1. Introduction

Deviations from perfect spherical symmetry in an ICF capsule's pressure drive can arise from several different sources. Non-uniformities in laser illumination due to variations in individual beam profiles and time integrated energies, beam pointing errors and target offsets from the nominal position of the beam focus in the target chamber can all lead to asymmetry in the pressure drive on the capsule. The impact of this asymmetry on capsule performance has been studied for a number of years because of its potential importance for ICF ignition attempts. Two recent experimental studies are of particular interest.

In Rygg *et al.*[1] the authors performed experiments with spherical plastic (CH) capsules with a $D^3He$ gas fill on the OMEGA laser system. Drive asymmetries were induced by offsetting the capsule from the nominal beam focus and the resulting capsule performance was measured. In this study the authors make the important distinction between the "shock burn" induced by the heating of the gas as a result of the initial collapse of the spherically converging gas shock on the capsule center and the subsequent "compression burn" induced by the compression and heating of the gas by the imploding shell. Results of this study demonstrate that the "shock burn" is relatively insensitive to the drive asymmetry. In contrast the "compression burn", which is the main component of the burn yield, is substantially diminished by increasing drive asymmetry.

In Li *et al.*[2] the authors again performed experiments with CH capsules on OMEGA using charged particle spectrometry techniques to measure the amplitude of low mode number $\rho R$ asymmetries in the imploded shell at the time of fusion burn and to show quantitatively how these low mode $\rho R$ asymmetries are directly correlated with the amplitude of asymmetries in the time-averaged, on-target laser intensity. From these and related studies the authors concluded that the observed growth of low mode $\rho R$ asymmetries in the imploded shell are initially seeded by laser illumination asymmetries and grow predominantly by Bell-Plesset[3] related convergence effects while capsule imperfections do not seem to play a dominant role in determining the observed $\rho R$ asymmetries.

In addition to results of these specific and very illuminating experiments, a closely related fundamental theoretical question arises: precisely how is turbulence generated in an ICF implosion and what, if any, is the relationship between drive asymmetry and the generation of turbulence in an ICF capsule? In this paper we present results from some very high resolution three-dimensional (3D) numerical simulations of the implosion of a highly idealized OMEGA capsule similar to the one used in the experiments of Rygg *et al.*[1], which address both the results of the above experiments and the more general question of how drive asymmetry can lead to turbulence generation in ICF implosions (cf. Thomas and Kares[4]).

Scaling studies are also presented which suggest the important role convergence may play in the hydrodynamic disruption of asymmetrically driven ICF implosions. These studies provide a qualitative understanding of the empirical observation that high convergence ICF implosions perform much worse in reality than would be expected from simulations based upon the unstable Rayleigh-Taylor growth of capsule surface irregularities alone. The prime example of this discrepancy at high convergence is the recent failure of the National Ignition Campaign (NIC) at the National Ignition Facility (NIF).

The simulations described here were performed using the Los Alamos National Laboratory's Eulerian Adaptive Mesh Resolution (AMR) radiation-hydrodynamics code RAGE[5] and utilize spatial resolutions down to 0.05 µm in 3D which approach the order of the Kolmogorov length scale in this system. While these inviscid simulations do not explicitly resolve the Kolmogorov length scale and are thus not direct numerical simulations (DNS) of the capsule implosion, we will show that the resolutions achieved are sufficient to see important details of how the turbulence is generated in an asymmetrically driven ICF implosion.

In particular we will see that the asymmetric pressure drive creates initial density perturbations in the drive shell that are amplified by Bell-Plesset related convergence effects. When the incoming shell collides with the outgoing reflected gas shock these perturbations act as seeds for the growth of Richtmyer-Meshkov fingers of shell material which penetrate the DT gas. The growth of these fingers is further enhanced at late time by the Rayleigh-Taylor instability. These same pressure drive asymmetries also create coherent vortical structures in the DT gas, counter-rotating rings and sheets of azimuthal vorticity of opposite sign in close proximity to one another. These structures are strongly unstable to both short and long wavelength azimuthal instabilities of the Widnall[6] and Crow[7] type. They very rapidly evolve to a fully turbulent state by a process similar to that described by Leweke and Williamson[8]. Thus, both the asymmetries in the shell and the turbulence in the gas are seen to have a common origin in the asymmetry of the pressure drive.

In Section 2 of this paper we discuss high resolution 2D and 3D RAGE simulations of a simplified OMEGA capsule implosion that illustrate the central role played by drive asymmetry in the determination of both the penetration of fingers of shell material into the gas and generation of the gas turbulence. In this Section we present linked 3D RAGE simulations of the turbulence generation process in the capsule implosion that use spatial resolutions as high as 0.05 µm and that require nearly $1 \times 10^9$ AMR cells in 3D. We demonstrate directly that the power spectra of the kinetic energy of the resulting turbulence observed in these simulations contains an inertial subrange whose scaling is consistent with a Kolmogorov $k^{-5/3}$ spectrum. Section 3 discusses possible implications of these results for NIF ignition. And finally, Section 4 summarizes our conclusions.

## 2. RAGE Simulations of an ICF Capsule with an Imposed Asymmetry

In this section we utilize high resolution 2D and 3D RAGE simulations to demonstrate the effect of imposing a well-defined drive asymmetry on an ICF capsule implosion. The example capsule chosen for our RAGE simulations is a plastic (CH) capsule similar to ones used in the experiments of Rygg *et al.*[1] on OMEGA. This capsule consisted of a round plastic shell with outer radius 425 µm and inner radius 400 µm containing an equimolar mixture of DT gas at an initial density of $2.5 \times 10^{-3}$ g/cm$^3$.

The goal of these simulations is not to model the results of an actual OMEGA experiment, but instead to use the plastic OMEGA capsule as a simple example, a perfectly spherical CH shell with DT gas, with which to explicitly exhibit how an asymmetric pressure drive leads to both the penetration of fingers of shell material into the DT fuel and the generation of turbulence in the DT fuel as the capsule implodes.

For this purpose we need only purely hydrodynamic calculations with no radiation or nuclear burn physics included and even these calculations are highly idealized. Rather than attempting to drive the capsule with an asymmetric radiation field, the energy was imparted to the plastic shell in the following way. The shell was divided into two regions, an inner region (the pusher) between $r = 400$ μm and $405$ μm and an outer region (the ablator) between $r = 405$ μm and $425$ μm. Energy was then sourced into this outer ablator region as an energy source per unit mass $S$ with a fixed spatial profile of the form,

$$S = At(1 + a_\ell P_\ell(\cos\theta))\tanh((r - r_0)/\Delta) \quad (1)$$

for $r > r_0$ and $0 < t < 1$ ns

where $r_0 = 405$ μm and $\Delta = 20$ μm. Here the value of $A = 1 \times 10^{16}$ ergs/gm/ns was chosen to source in a total of 17.35 kJ in a 1 ns square power pulse. This form for the energy deposition was chosen to suppress grid-induced Rayleigh-Taylor (RT) instabilities at the ablator/pusher interface driven by the acceleration of the ablator upon the denser pusher. This treatment allows the simulation to retain only the imprinting from the drive asymmetry without having to worry about possible feed-through of unwanted short wavelength RT instabilities. The pusher region moves under the influence of the drive pressure generated in the ablator. No energy was sourced directly into the pusher so that a high density pusher region was present late in the implosion.

The term proportional to $P_\ell(\cos\theta)$, the Legendre polynomial of order $\ell$, provided a simple asymmetry of amplitude $a_\ell$ for the drive. The results presented here focus on the case of an energy source with an $\ell = 30$ asymmetry. A drive with P30 asymmetry was chosen so that a rather large number of features is created, which may have some relevance to a laser drive with a large number of beams. However, no attempt has been made to use real asymmetries from actual laser drives but rather we have chosen a simple asymmetry that clearly illustrates the physical phenomena associated with asymmetric drive.

In this paper we are interested in examining the fundamental mechanisms of turbulence production in our example ICF implosion, a study which requires very high resolution 3D numerical simulations of the implosion. However, such simulations are so computationally intensive that they cannot be practically carried out from $t = 0$ with the current generation of parallel supercomputers. To avoid this difficulty we have adopted the procedure of performing the initial phase of the implosion in 2D using the RAGE code. Then at a suitably chosen time late in the implosion the 2D RAGE simulation is rotated in 3D and used to initialize a 3D RAGE simulation of the late time behavior of the implosion. This 2D to 3D linking procedure allows us to carry out very high resolution simulations at late time of the turbulence generation while avoiding the practical limitations of existing compute platforms. However, it also restricts us to examining only axisymmetric rather than fully three-dimensional asymmetries of the pressure drive. Despite this restriction we shall see that the basic mechanisms of turbulence generation are clearly exhibited. We now turn to results from 2D RAGE simulations of our idealized capsule implosion.

### 2.1 2D RAGE Simulations of the Capsule Implosion

Fig. 1 shows an r-t plot of the motion of the pusher/DT gas interface (solid magenta curve) from a 1D RAGE simulation of our idealized OMEGA capsule. This plot shows the abrupt acceleration, the roughly constant velocity implosion and the stagnation phase with the multiple shock bounces. The solid orange curve shows the corresponding r-t motion for the main gas shock, which suffers its first collapse onto the capsule center at about $t = 1.08$ ns. We are primarily interested in the implosion up to the stagnation point at about $t = 1.75$ ns, since beyond this time in a real experiment the burn would have begun. The convergence ratio for the initial to final radius of the pusher/gas interface at $t = 1.75$ ns is 7.9 in this 1D simulation. Superimposed in Fig. 1 on the r-t plot of the pusher/gas interface from the 1D RAGE simulation is the corresponding result from a 2D RAGE simulation (hollow blue squares) for the case of a spherically symmetric drive with $a_{30} = 0$. The interface motion in the symmetrically driven 2D RAGE simulation tracks the interface motion in the 1D simulation extremely well.

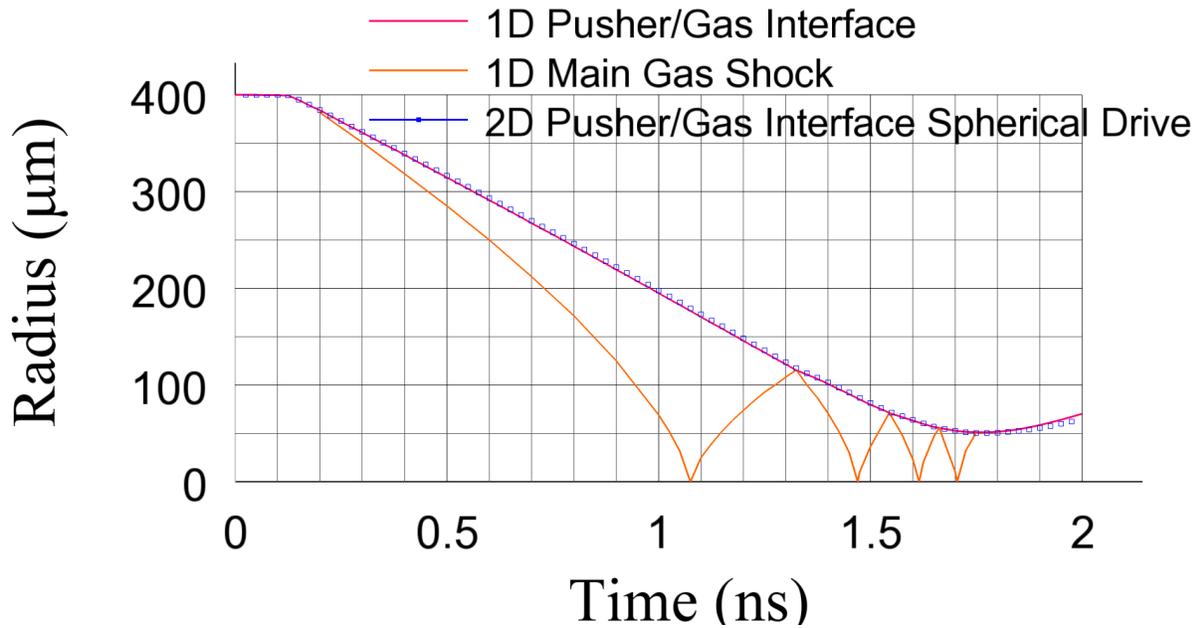

**Fig. 1.** r - t plot of the motion of the pusher/gas interface and the main gas shock in a 1D RAGE simulation of the OMEGA capsule. Blue squares show the pusher/gas interface motion in a 2D RAGE simulation using spherically symmetric drive with $a_{30} = 0$. Effective interface radius in 2D is computed from the DT gas volume using $R_{eff} = (3 \cdot V_{gas} / 4\pi)^{1/3}$.

2D RAGE simulations were performed with a range of different values for the amplitude $a_{30}$ of the P30 drive asymmetry from $a_{30} = 0$ (spherically symmetric drive) to $a_{30} = 0.50$ (50% P30 asymmetry) with the same total energy of 17.35 kJ. All of these 2D RAGE simulations had a nominal convergence of about 8 for the pusher/gas interface. The 2D simulations use adaptive mesh refinement with a finest spatial resolution of 0.4 μm up until a time of $t = 1$ ns just before the first collapse of the main gas shock on the center of the capsule at around $t = 1.08$ ns. After that time a fixed spatial resolution of approximately 0.4 μm is forced throughout the entire region that encompasses the pusher and the gas in order to achieve the highest fidelity for the dynamics of the gas and the pusher shell during the later stages of the implosion.

Fig. 2 shows a series of panels, each a time snapshot from the early portion of a 2D RAGE simulation of the example OMEGA capsule for the case with a 50% P30 asymmetry which shows how the implosion progresses early in time as the asymmetric drive is applied. In the lower portion of each panel the CH shell which includes both the ablator and pusher regions is colored by pressure while the DT gas is colored by grad P. In the upper portion of each panel both the CH shell and the gas are colored by the θ component of the material velocity. The panels are arranged in rows with the left-hand panel in each row showing a full view of the capsule at the particular snapshot time while the corresponding right-hand panel shows a blowup of a small region at the left pole of the capsule near the CH/gas interface at that same time.

Fig. 2(a) at $t = 0.1$ ns shows the angular position of the 16 pressure maxima associated with the P30 drive. The drive produces predominantly radial flow, but also produces flow in the θ direction away from the maxima as indicated in the figure because the drive is not perfectly spherically symmetric. At this early time of $t = 0.1$ ns, motion is confined to the ablator shell.

By $t = 0.275$ ns (Figs. 2(c) and 2(d)) the main shock has begun to penetrate into the gas. Behind this inward moving main shock, gas is flowing in the θ direction away from the angular positions of the original drive pressure maxima in exactly the same way as the adjacent shell material. While the main shock moves inward, the gas flow behind the main shock converges in the θ direction at the angular positions of the original drive pressure *minima*. Gas collides there forming pairs of shocks behind the main shock which spread out in the θ direction away from the angular position of the original drive minima. Fig. 2(d) at $t = 0.275$ ns shows these polar shocks just beginning to form behind the main shock as gas collides at the angular position of the original drive minima.

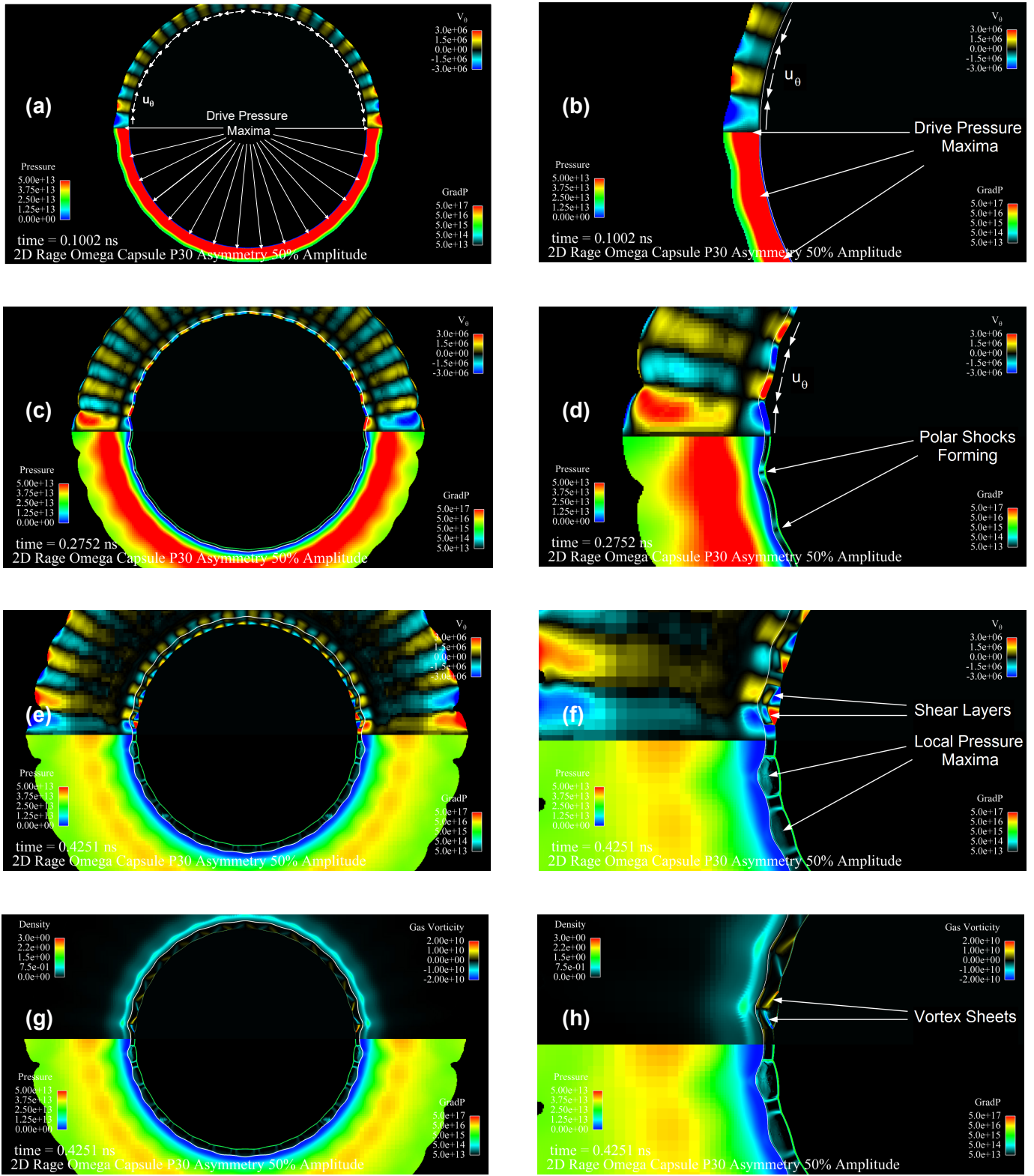

**Fig. 2.** A sequence of early time snapshots from a 2D RAGE simulation of the P30 Omega capsule with $a_{30} = 0.50$. In the lower portion of each panel the CH shell is colored by pressure while the gas is colored by grad P. In the upper portion both the CH shell and the gas are colored by the $\theta$ component of the material velocity. In the last row of panels the $\theta$ velocity has been replaced on top by density in the CH and azimuthal vorticity in the gas. Each right hand panel shows an expanded view of the region at the left pole of the capsule near the CH/gas interface.

As a result of this convergent flow in the θ direction, local pressure maxima form in the gas at the angular positions of the original drive minima. These local pressure maxima produce additional grad P accelerations directed away from the local pressure maxima which drive gas inward into the unshocked region. By $t = 0.425$ ns (Figs. 2(e) and 2(f)) these local pressure maxima have produced convex perturbations on the main shock as well as regions of gas behind the main shock in which the θ component of the gas velocity has reversed sign. The result is the formation of shear layers in the gas which can be clearly seen in the θ velocity field shown in the upper portion of Fig. 2(f) at $t = 0.425$ ns.

In Figs. 2(g) and 2(h), also at the same time of $t = 0.425$ ns, we have replaced the θ velocity in the upper portion of the panel by density in the CH shell and the azimuthal component of vorticity in the gas. This pair of panels illustrates that the shear layers seen in the θ velocity field of Figs. 2(e) and 2(f) correspond to vortex sheets in the gas. As the main shock travels inward, the moving intersection points between the main shock and the polar shocks formed behind it trace out these vortex sheets in the gas. This process is particularly evident in the expanded view in Fig. 2(h) at $t = 0.425$ ns. Notice that as a result of this mechanism, vortex sheets of opposite sign for the azimuthal vorticity, yellow for positive and blue for negative, always form in close proximity in the gas.

This same asymmetry of the pressure drive also produces convergent flow in the θ direction inside the shell material as well. The result are the 15 density maxima in the shell at the angular positions of the original drive minima which are clearly visible in Figs. 2(g) and 2(h) at $t = 0.425$ ns. As the capsule implodes these initial density perturbations seeded by the asymmetry of the pressure drive are amplified by Bell-Plesset related convergence effects.

Fig. 3 shows a series of panels, each a time snapshot from the remainder of the implosion. Here in the lower portion of each panel the CH shell is colored by pressure while the gas is colored by grad P. In the upper portion of each panel the CH shell is colored by density while the gas is colored by the azimuthal component of the gas vorticity. Fig. 3(a) at $t = 0.6$ ns shows that as the polar shocks pass through each other new layers of vortex sheets are formed in the gas by the process described above.

This process continues as the main shock converges radially inward toward the center of the capsule so that by $t = 1.025$ ns (Fig. 3(b)), a time just before the first bounce of the main shock, several layers of vortex sheets have been formed within the gas. By $t = 1.1$ ns (Fig. 3(c)) the main shock has bounced off the center of the capsule and is moving radially outward. As this outbound main shock passes over the vortex sheets, it rolls them up into vortex rings, a process that is clearly visible in the $t = 1.225$ ns panel in Fig. 3(d).

Simultaneous with the deposition and evolution of the vorticity in the gas is the development of pressure and density enhancements in the plastic shell as can been seen in the top four panels of Fig. 3. Because the original pressure drive was not spherically symmetric, the shell material flows in the θ direction as shown in Fig. 2(a) and deterministic pressure and density enhancements develop in the shell at the angular positions of the original drive minima. These perturbations grow in time due to convergence (the Bell-Plesset effect) becoming increasingly elongated in the radial direction until shell material protrudes out into the gas as shown in the $t = 1.225$ ns panel in Fig. 3(d). At a slightly later time $t = 1.325$ ns (Fig. 3(e)) the outgoing reflected main gas shock collides with the shell/gas interface and these perturbations are amplified by the Richtmyer-Meshkov instability to create large scale deterministic fingers of shell material in the gas. The continued growth of these deterministic Richtmyer-Meshkov fingers is seen in the last three panels of Fig. 3.

The collision of the outgoing gas shock with the perturbed shell/gas interface also deposits significant vorticity on the interface as part of the Richtmyer-Meshkov development. For a compressible, inviscid

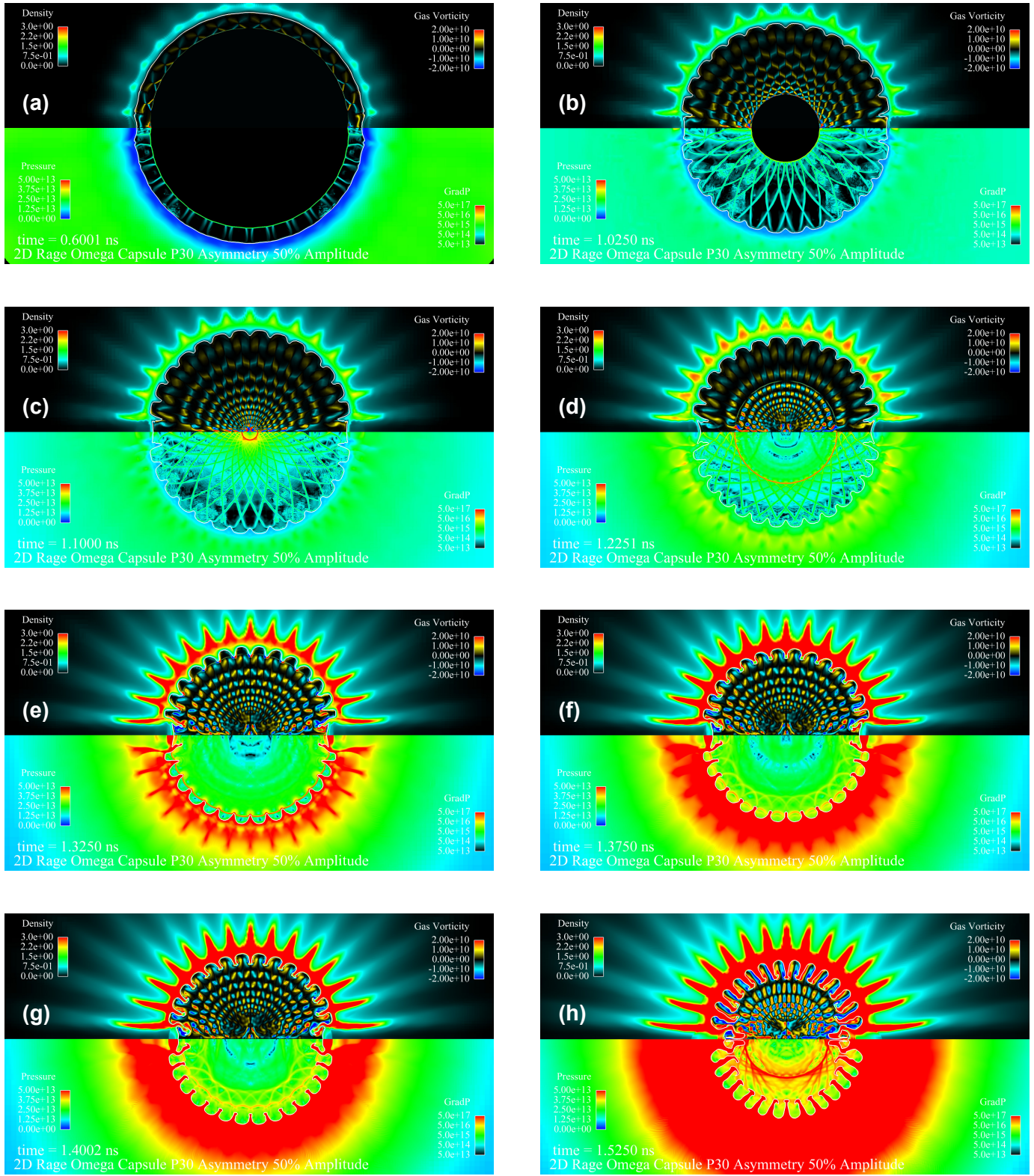

**Fig. 3.** Late time snapshots from 2D RAGE simulation of the P30 Omega capsule with $a_{30} = 0.50$. In the lower portion of each panel the CH shell is colored by pressure while the gas is colored by grad P. In the upper portion the CH shell is colored by density while the gas is colored by azimuthal vorticity.

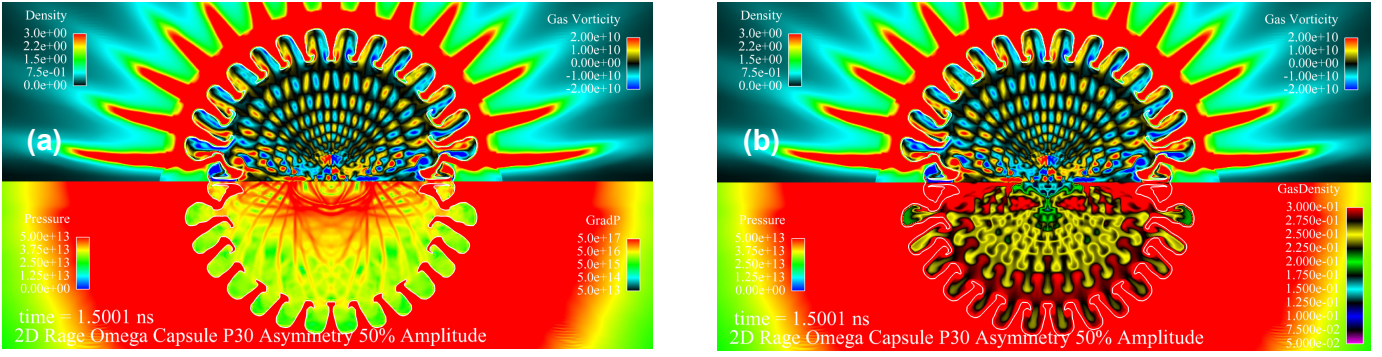

**Fig. 4.** Snapshots at a time *t* = 1.50 ns from the 2D RAGE simulation of the idealized OMEGA capsule with $a_{30}$ = 0.50 . The gas region in the lower portion of the left-hand panel is colored by the pressure gradient. The gas region in the lower portion of the right-hand panel is colored by the gas density. Note the complex density structure in the interior of the gas which corresponds to the gas azimuthal vorticity in the upper portion of the right hand panel. The time shown is just after the second bounce of the gas shock off the capsule center which occurred at *t* = 1.485 ns.

simulation, vorticity production is controlled by the vorticity evolution equation,

$$\frac{\partial \vec{\omega}}{\partial t} + (\vec{v} \cdot \nabla)\vec{\omega} = \frac{1}{\rho^2}\nabla\rho \times \nabla P - \vec{\omega}(\nabla \cdot \vec{v}) + (\vec{\omega} \cdot \nabla)\vec{v} \quad (2)$$

where ρ and *P* are the scalar density and pressure and where,

$$\vec{\omega} = \nabla \times \vec{v}$$

is the vorticity of the velocity field $\vec{v}$ in the simulation. The second term on the right in Eqn. (2) arises from the compressibility of the fluid. The third term on the right in Eqn. (2) is the vortex stretching term which for axisymmetric 2D simulations is identically zero. The first term on the right in Eqn. (2), the baroclinic torque term, results in vorticity production wherever pressure and density gradients are not aligned. In particular, where the outgoing reflected gas shock in our 2D asymmetric capsule simulation collides at an oblique angle with the perturbed surface of the shell/gas interface, then vorticity is deposited along the interface as a result of the baroclinic torque produced.

This process can be observed in the *t* = 1.325 ns snapshot of Fig. 3(e) where red and blue vortex sheets of opposite sign for the azimuthal vorticity can been seen forming along opposite sides of the fingers of shell material as the outgoing reflected gas shock sweeps radially outward over them. These vortex sheets are a result of the baroclinic torque generated by the oblique collision of the gas shock with the perturbed surface of the shell/gas interface. As the large scale fingers develop, much of this vorticity is rolled up into vortex rings of opposite sign which are trapped in the developing bubbles of gas between the growing Richtmyer-Meshkov mushrooms. By the end of the 2D RAGE simulation, at *t* = 1.525 ns shown in Fig. 3(h), there are large scale fingers of shell material in the gas as well as red and blue counter-rotating vortex rings in both the body of the gas and in the developing bubbles of gas between the growing Richtmyer-Meshkov fingers. Such counter-rotating ring pairs in close proximity are typically highly unstable in 3D.

Fig. 4 illustrates the complex structure in the gas density associated with these counter-rotating vortex rings. Fig. 4(a) shows another snapshot taken from the time sequence in Fig. 3 at a time *t* = 1.50 ns just after the second bounce of the main gas shock off the capsule center. In the lower portion of Fig. 4(a) the gas is colored by the pressure gradient to reveal the complex reflected gas shock traveling radially outward. In the lower portion of Fig. 4(b) the gas is colored by density instead to reveal the intricate structure of density mushrooms in the interior of the gas associated with the counter-rotating vortex rings there. Fig. 4(b) serves to emphasize the complex dynamical nature of the gas which arises from the non-radial flow in the gas due to drive asymmetry.

Fig. 5 compares results from six separate 2D RAGE simulations of the OMEGA capsule implosion in which the amplitude of the imposed P30 drive asymmetry is varied with a fixed total drive energy of 17.35 kJ. Simulation results for $a_{30}$ values of 0.00 (spherically symmetric), 0.01, 0.05, 0.10, 0.25 and 0.50 are shown which illustrate the effect of increasing the amplitude of the drive asymmetry. Fig. 5 shows time snapshots from each of these six simulations at a time *t* = 1.55 ns shortly before stagnation. The convergence ratio of the pusher/gas interface is 5.6 for the spherically symmetric case at this time. Fig. 5(a) clearly shows that for this choice of drive energy and a resulting convergence ratio of 5.6, the symmetrically driven RAGE simulation with $a_{30}$ = 0 gives a round shell with only minor numerical artifacts apparent at the 45° diagonal directions as well as a round main gas shock, even after two bounces of the main shock off the center of the capsule. In contrast, the 2D RAGE simulation shown in Fig. 5(f), which was driven asymmetrically with a P30 component of amplitude 50% ($a_{30}$ = 0.50), shows a great deal of real structure which is due to the imposed drive asymmetry. Fifteen well-defined Richtmyer-Meshkov fingers of shell material are visible in this bottom right hand panel which are growing radially inward together with a pattern of strong counter-rotating vortices which have been generated in the gas. The snapshots in Fig. 5 illustrate how such structures continuously emerge as the amplitude of the P30 drive asymmetry is increased from 0% to 50%. Note, for example, how the 15 finger structure of the gas region which is the result of the P30 asymmetry is evident in all the implosions with a non-zero value of $a_{30}$ , even in the 1% amplitude case shown in Fig. 5(b) where a slight 15-fold perturbation visible on the CH/gas interface reveals the action of the P30 drive asymmetry.

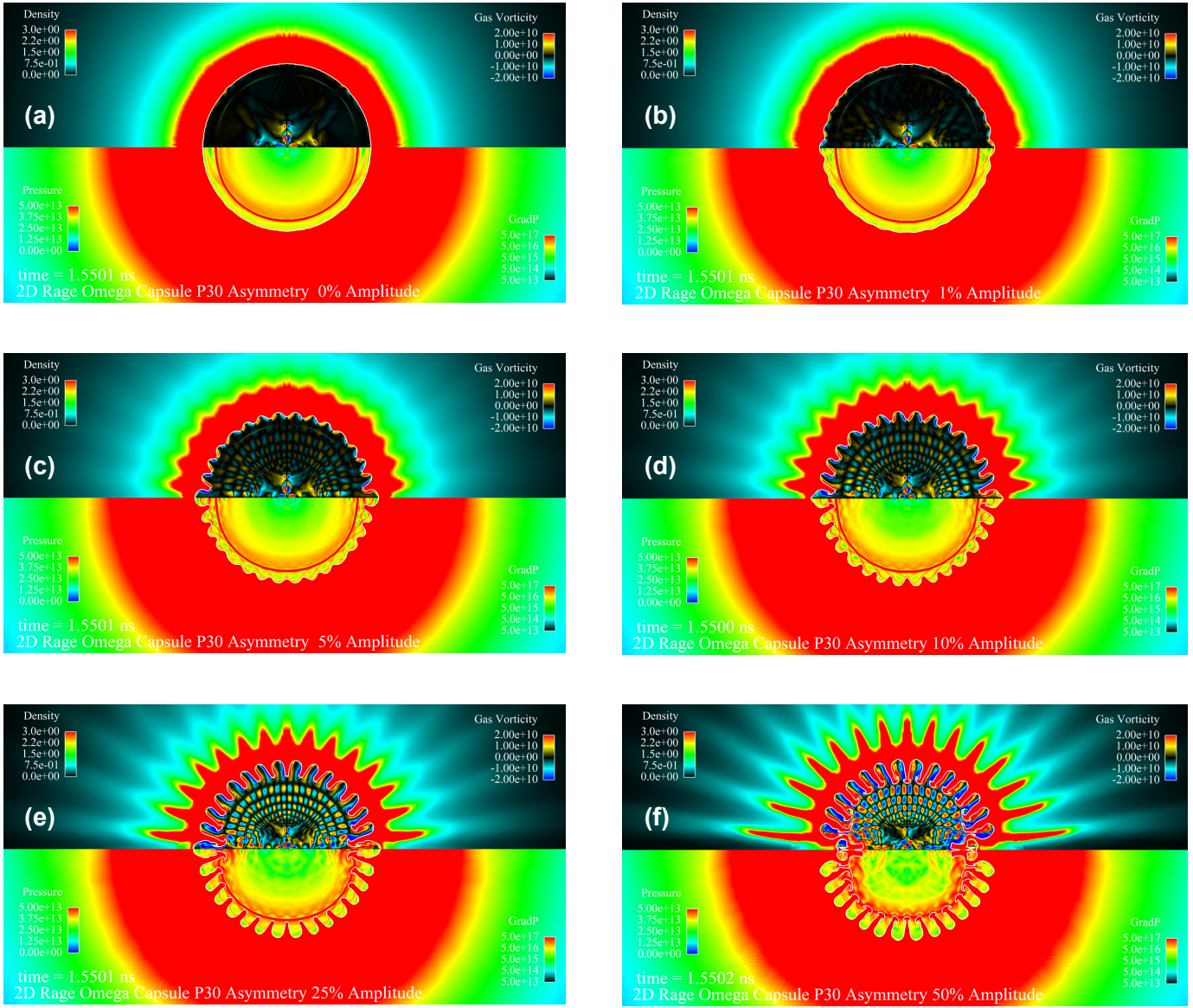

**Fig. 5.** Snapshots at a fixed time *t* = 1.55 ns from six separate 2D RAGE simulations of the idealized OMEGA capsule illustrating the effect of varying the amplitude of the P30 asymmetry. Simulation results for $a_{30}$ values of 0.00 (spherically symmetric), 0.01, 0.05, 0.10, 0.25 and 0.50 are shown. The time of these snapshots is chosen to be about 0.20 ns before minimum gas volume in the $a_{30}$ = 0 simulation at *t* = 1.75 ns. In the lower portion of each panel the CH shell is colored by pressure while the gas is colored by grad P. In the upper portion the CH shell is colored by density while the gas is colored by azimuthal vorticity.

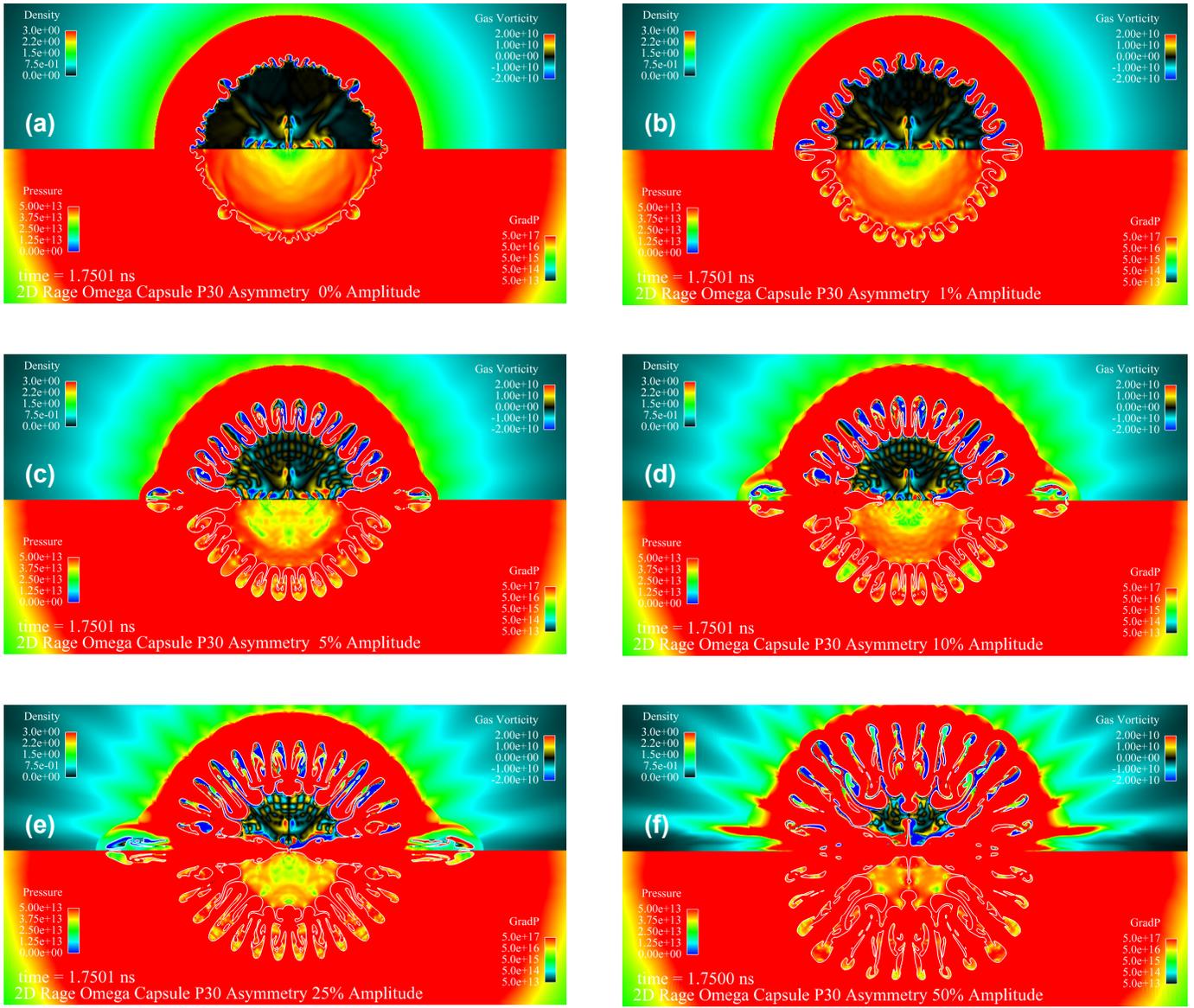

**Fig. 6.** Snapshots at a later time *t* = 1.75 ns from the same six 2D RAGE simulations of the idealized OMEGA capsule. The time of these snapshots is chosen to be near minimum gas volume so that all of the implosions have stagnated at this time. Note the appearance, as expected, of irregular Rayleigh-Taylor structure on the CH/gas interface for the case in panel (a) with spherically symmetric drive. Note also that even with an imposed P30 asymmetry of only 1%, a well-defined pattern of 15 fingers of shell material associated with the P30 asymmetry is clearly visible in panel (b).

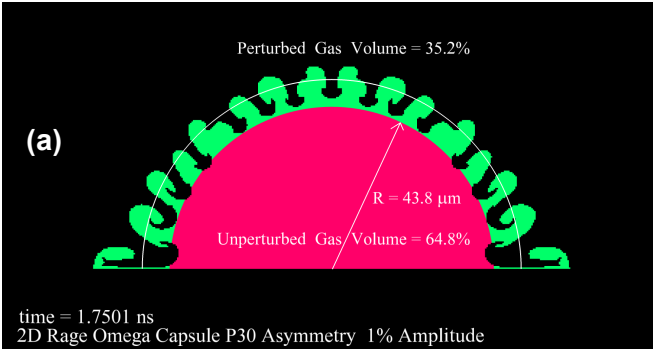 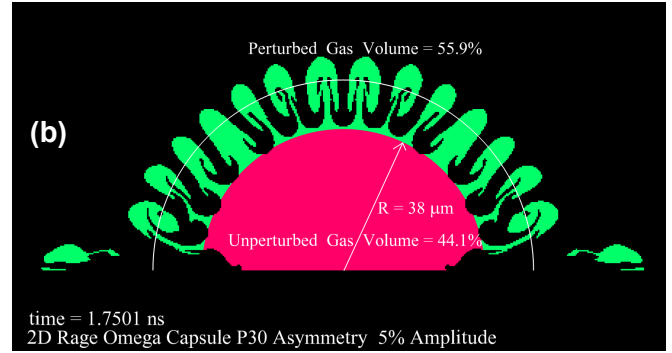

**Fig. 7. Snapshots at a time *t* = 1.750 ns from the two 2D RAGE simulations of the idealized OMEGA capsule with $a_{30}$ = 0.01 and 0.05 . The gas has been divided into two regions, a central unperturbed core and a region perturbed by the growth of fingers of shell material into the gas. For 1% imposed asymmetry (a) the perturbed region contains 35.2 % of the total gas volume. For 5% imposed asymmetry (b) the perturbed region contains 55.9% of the total gas volume. The white circle of radius 51.17 μm represents the position of the CH/gas interface in the 1D RAGE simulation.**

Fig. 6 shows the same six snapshots at a later time *t* = 1.75 ns corresponding to the stagnation of the implosion. Here the spherically symmetric case in Fig. 6(a) shows the development of irregular Rayleigh-Taylor structure at the CH/gas interface which grows from initial numerical perturbations provided by the grid as would be expected in this physically RT unstable situation. Compare this, however, with Fig. 6(b) which shows the $a_{30}$ = 0.01 case at this same time. Even at a 1% amplitude for the P30 asymmetry the 15 fingers of shell material which arise from the P30 asymmetry are clearly evident. For amplitudes of 5%, 10%, 25% and 50% the P30 asymmetry overwhelmingly dominates the final structure of the gas region. As Fig. 7(a) illustrates, even in the 1% case, at stagnation the fingers of shell material extend inward about 14% of the 51.17 μm radius of the gas region obtained in the 1D RAGE simulation, which corresponds to about 35% of the gas volume being significantly perturbed. For the 5% asymmetric drive (Fig. 7(b)) the fingers penetrate more than 26% of the radius of the gas region and about 56% of the total gas volume is perturbed at stagnation. For larger asymmetries the vast majority of the gas volume is severely perturbed by the presence of the fingers at stagnation. These fingers are at well-defined angular positions since their geometry is basically determined by the drive asymmetry followed by growth due to convergence and, later in time, the Richtmyer-Meshkov instability. Fingers of this type may be related to the angular asymmetries in ρR inferred in the proton spectrometry experiments of Li et al.[2]. With the proper choice of asymmetric drive the shell material fingers might be created in specific locations to test how well one might be able to predict such behavior.

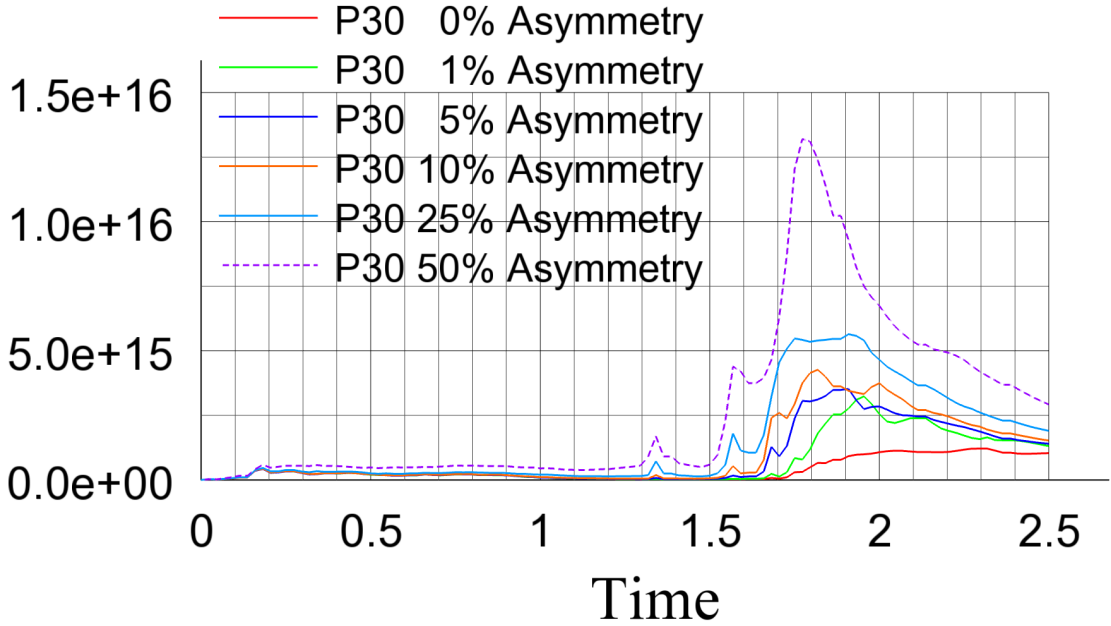

**Fig. 8. Time history for the total enstrophy for each of the six 2D OMEGA simulations of Fig. 4 with imposed P30 drive asymmetries of 0, 1, 5, 10, 25 and 50%. Enstrophy is in units of cm³/sec².**

Another numerical artifact which is most clearly visible for the spherically driven case with $a_{30} = 0$ in Fig. 5(a) is the spurious vorticity generated at the center of the capsule by the spherical collapse of the gas shock. This spurious vorticity cannot be eliminated if a fixed Eulerian square grid is used to compute the spherical collapse of the gas shock onto the capsule center, as in these 2D RAGE simulations. As would be expected, a similar vorticity artifact is seen at the capsule center in all of the panels of Fig. 5 for all values of $a_{30}$ from 0 to 50%. However, because this spurious vorticity occurs at small radius it makes a negligible contribution to the total enstrophy of the simulation. The total enstrophy of one of the 2D simulations can be computed as,

$$\int \tfrac{1}{2}|\omega_\phi|^2 \, 2\pi r \, dr \, dz \qquad (3)$$

where $\omega_\phi$ is the azimuthal component of vorticity in cylindrical coordinates $(z, r, \phi)$ and the integral is extended over the entire axisymmetric 2D simulation volume.

Fig. 8 shows time history plots of the total enstrophy for all six of the 2D RAGE simulations. Early in time before $t = 1$ ns there is a small but non-zero contribution to the total enstrophy from the region outside of the capsule as a result of the expansion of the ablated plastic against the background gas. However, by about $t = 1$ ns this contribution has largely dissipated and for the calculations with low drive asymmetry little total enstrophy is generated prior to the first collision of the outgoing reflected gas shock with the incoming CH/gas interface at about $t = 1.34$ ns. The increased total enstrophy for the 50% asymmetry case in the time period prior to $t = 1.34$ ns is primarily a result of the enstrophy associated with the strong vortex rings produced in the body of the gas as seen in Fig 3. Sharp increases in total enstrophy are seen in all simulations with non-zero values of $a_{30}$ at times near $t = 1.34$ ns, $t = 1.56$ ns and $t = 1.68$ ns corresponding to the first, second and third collisions of the outgoing reflected gas shock with the incoming CH/gas interface. Beyond this time the structure of the reflected shocks traversing the gas becomes increasingly complex and the observed increase in total enstrophy at late time arises from multiple complex gas shock interactions with the perturbed CH/gas interface combined with late time Rayleigh-Taylor growth. The CH/gas interface is more perturbed for larger asymmetries and so the total vorticity production is larger for larger asymmetries.

Of particular note is that the spherically driven simulation (red curve) in Fig. 8 shows a very low total enstrophy until after stagnation when the CH/gas interface is grossly Rayleigh-Taylor unstable. For the spherically driven case, no significant enstrophy generation is observed either from the collapse of the gas shock on axis or the collision of the reflected gas shock at the CH/gas interface. This observation provides quantitative evidence that the spherically driven 2D simulation remains spherically symmetric to a high degree and that the spurious vorticity seen near the origin in these 2D simulations plays only a small role in the overall behavior of the system.

We now turn to a discussion of some detailed scaling studies of our idealized OMEGA capsule performed using 2D RAGE. In these studies we consider variations of the basic P30 implosion that include changing the mode number of the drive asymmetry, amplification of ρR non-uniformities of the CH shell and the effect of changing the initial fill density of the DT gas and the consequent increase in the convergence ratio of the implosion. For these studies we have adopted a slightly different gridding strategy from the one used for the 2D simulations of the P30 capsule discussed above. Here we have utilized the AMR control of the mesh available in RAGE to force a very high spatial resolution of 0.05 μm in the region of the CH pusher shell at the expense of sharply reduced resolution in the body of the DT gas. The purpose of this gridding strategy is to focus computational resources on the dynamical behavior of the CH shell in 2D. We begin by considering the effect of varying the mode number of the asymmetric drive on the implosion.

### 2.1.1 Effect of Varying the Mode Number of the Imposed Asymmetry

Fig. 9 shows a series of time snapshots at $t = 1.75$ ns, the time of stagnation in the 1D implosion, from eight different 2D RAGE simulations of our idealized OMEGA capsule implosion with varying $\ell$ values for the imposed drive asymmetry. Fig. 9(a) shows the case with no imposed asymmetry, the spherically symmetric case. In Fig. 9(a) only random short wavelength Rayleigh-Taylor structure is visible on the inner surface of the shell as would be expected in this case. For the case of $\ell = 4$ shown in Fig. 9(b) the long wavelength low order perturbation is visible with superimposed random short wavelength Rayleigh-Taylor structure on the inner surface of the shell. Note, however, that as the order of the perturbation is increased from $\ell = 14$ to $\ell = 20$ (Figs. 9(e) though 9(g)), a coherent structure emerges on the inner surface of the shell containing $\ell/2$ large scale fingers of shell material that protrude into the DT gas. In the final snapshot of Fig. 9(h) with $\ell = 30$ the structure with 15 large fingers observed in our previous P30 simulations is again seen.

This result is consistent with the well-known elementary estimate (cf. Yabe [9]) that the most dangerous mode for the disruption of an imploding spherical shell with initial radius $R_0$ and thickness $\Delta R_0$ by Rayleigh-Taylor growth has a mode number of order, $\ell = R_0 / \Delta R_0$. For the current case with $R_0 = 412.5$ μm and $\Delta R_0 = 25$ μm this simple estimate gives $\ell = 16.5$. Thus, from elementary considerations we might expect for modes numbers around $\ell = 16$ to observe the growth of large amplitude penetrations of the DT gas by fingers of capsule material and that is indeed what is observed in the sequence of Fig. 9.

### 2.1.2 Growth and Scaling in the ρR Asymmetries of the CH Shell

A quantity of particular interest experimentally is the angular distribution in the ρR of the imploding shell which has been measured for OMEGA capsules in the proton spectrometry experiments of Li *et al.*[2] Here we present results for the angular distribution of ρR for the CH shell obtained from a 2D RAGE simulation of our idealized P30 capsule with the nominal gas fill density of $\rho_i = 2.5 \times 10^{-3}$ g/cm$^3$ and an imposed drive asymmetry of 5%, $a_{30} = 0.05$. The effective ρR along any line of sight from the center of the capsule that passes completely through the imploding shell is defined as the line integral of the shell density along that line of sight. By varying the angular position of the line of sight at a fixed time in the simulation and computing the line integral of the shell density along that line of sight, we can generate the angular distribution for the ρR of the imploding shell. Once this distribution is obtained at each simulation time, both the angular average <ρR> and the rms deviation from this average <δρR> can also be computed as a function of time. Figs. 10(a) though (d) present snapshots from four different times in the 5% 2D RAGE simulation. The center of each snapshot shows a view of the capsule with the pusher region colored in gray and the interior of the fill gas colored by the gradient of pressure to illustrate the position of the gas shocks. The plotter in the lower left corner of each snapshot shows the angular distribution of ρR plotted versus cos(θ) at the corresponding simulation time. Also indicated on this plotter are the values of <ρR> and +/-<δρR> at that simulation time. The plotter in the lower right corner of each snapshot shows a time history plot of <δρR>/<ρR> with a gray vertical line indicating the current simulation time. The plotter in the middle right region of each snapshot shows a time history plot of <ρR> where, again, a gray vertical line indicates the current simulation time. The plotter in the upper right corner of each snapshot shows a time history of the convergence ratio of the CH/gas interface $C_r = R(t)/R_0$ where the effective interface radius in 2D is computed from the DT gas volume using $R(t) = (3 \cdot V_{gas} / 4\pi)^{1/3}$ and where again the current simulation time is indicated by the gray vertical line. The plotter in the upper left corner of each snapshot is a time history of the scaled fractional ρR asymmetry which we will discuss in greater detail shortly.

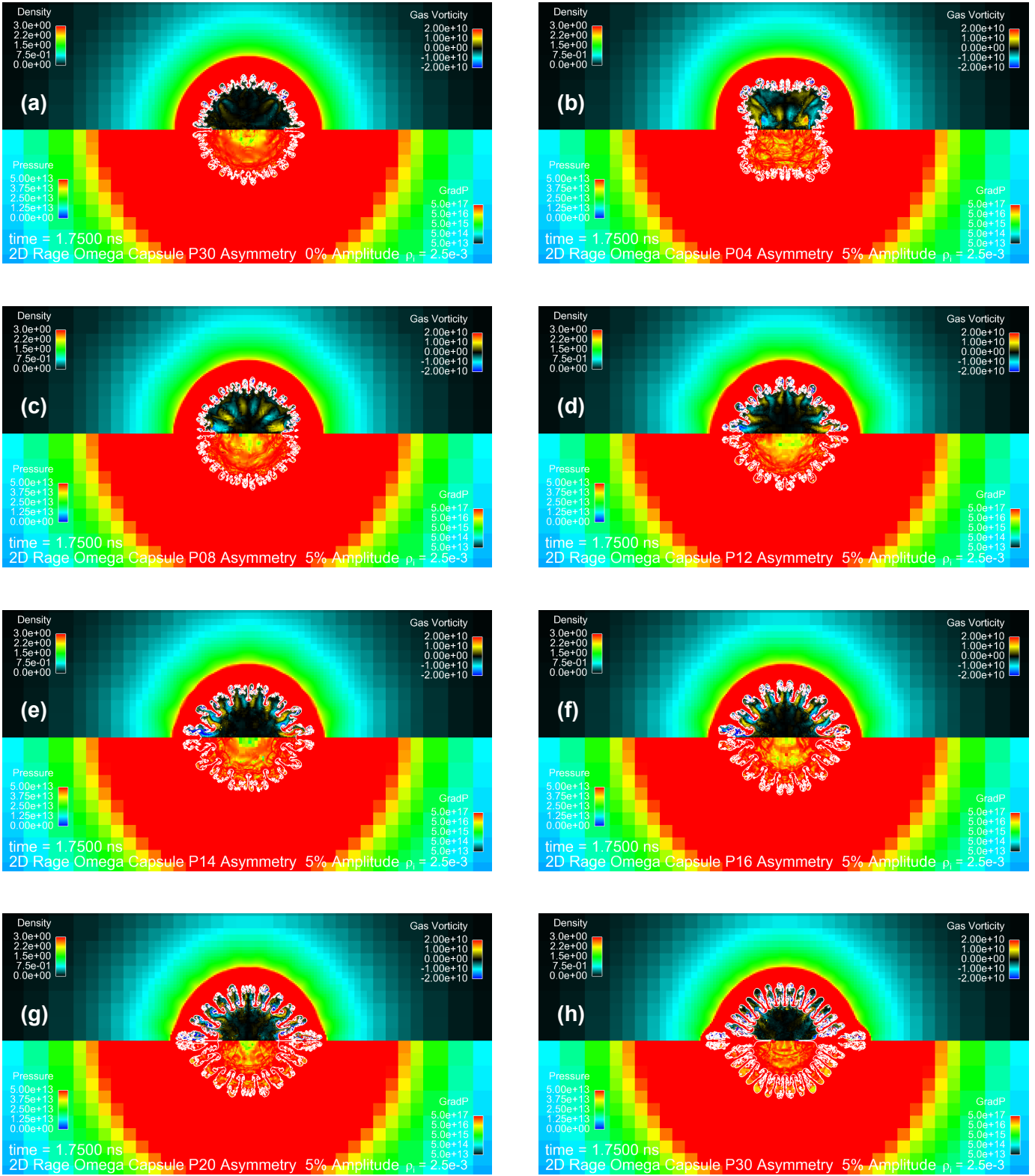

**Fig. 9.** Panels (b) – (h) show snapshots at *t* = 1.75 ns from seven different 2D RAGE simulations of the idealized OMEGA capsule with $a_\ell$ = 0.05 varying the mode number of the imposed drive asymmetry with $\ell$ = 4, 8, 12, 14, 16, 20 and 30. Panel (a) shows the case with no imposed asymmetry. Near $\ell$ = 16 shown in panel (f) a clear mode structure emerges on the inner surface of the shell with $\ell/2$ fingers of shell material that reflects the symmetry of the imposed perturbation.

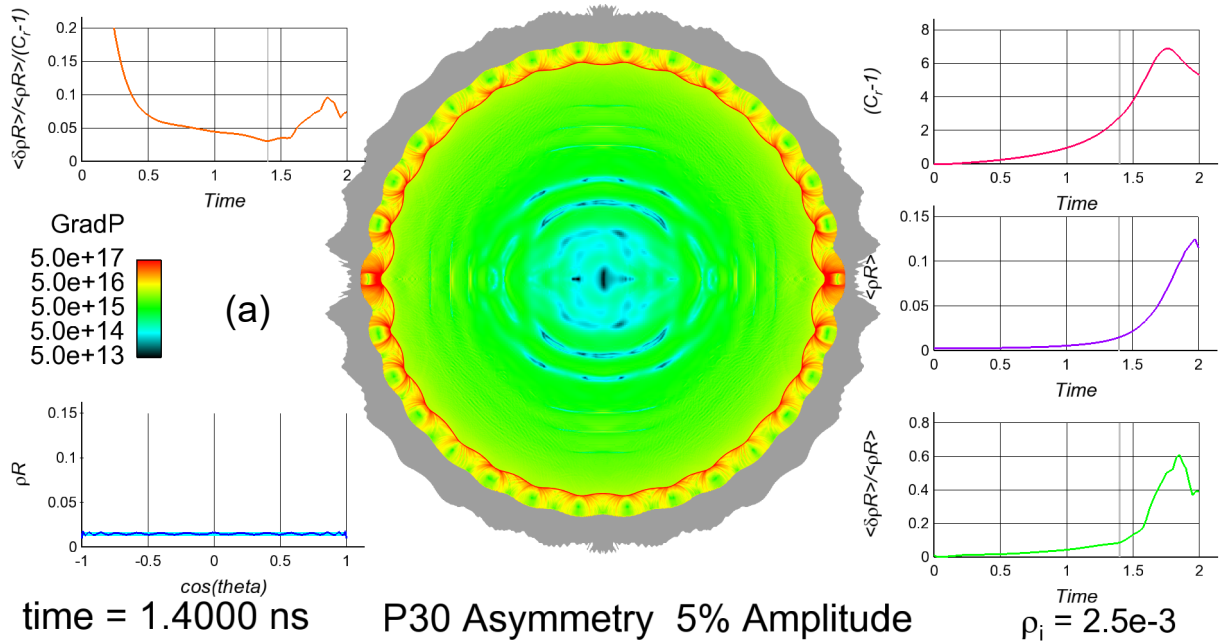

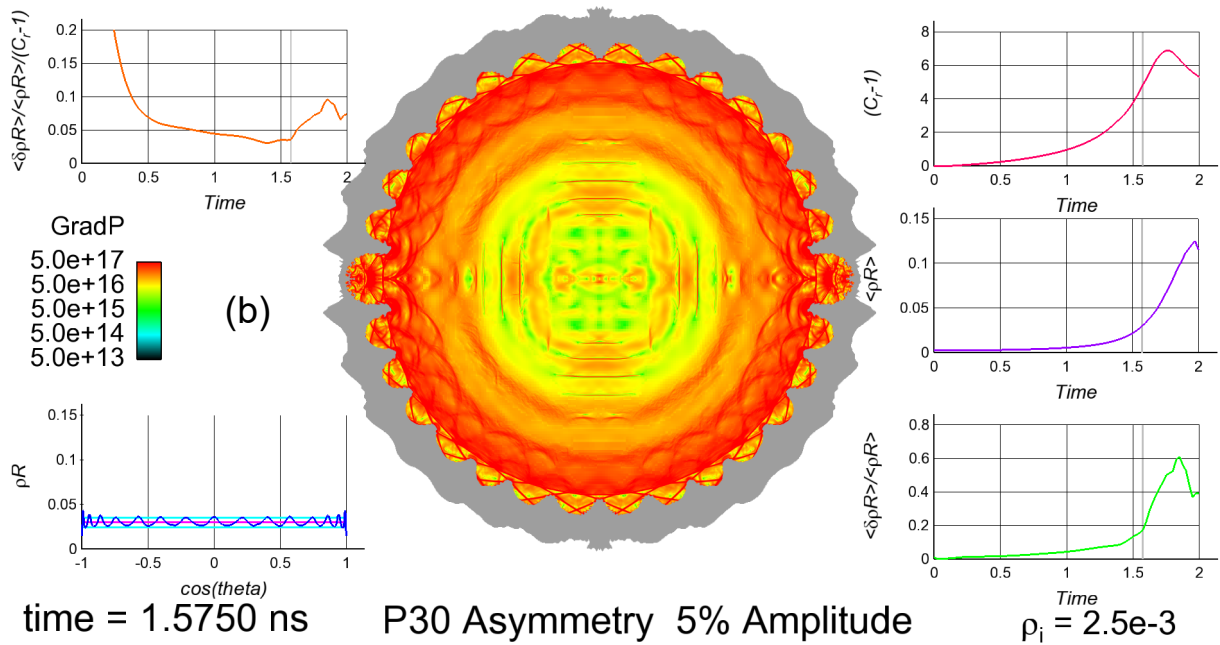

**Fig. 10. Four time snapshots from a 2D RAGE simulation of the P30 capsule with $a_{30}$ = 0.05 illustrating the rapid growth in $\rho$R asymmetries of the CH shell.**

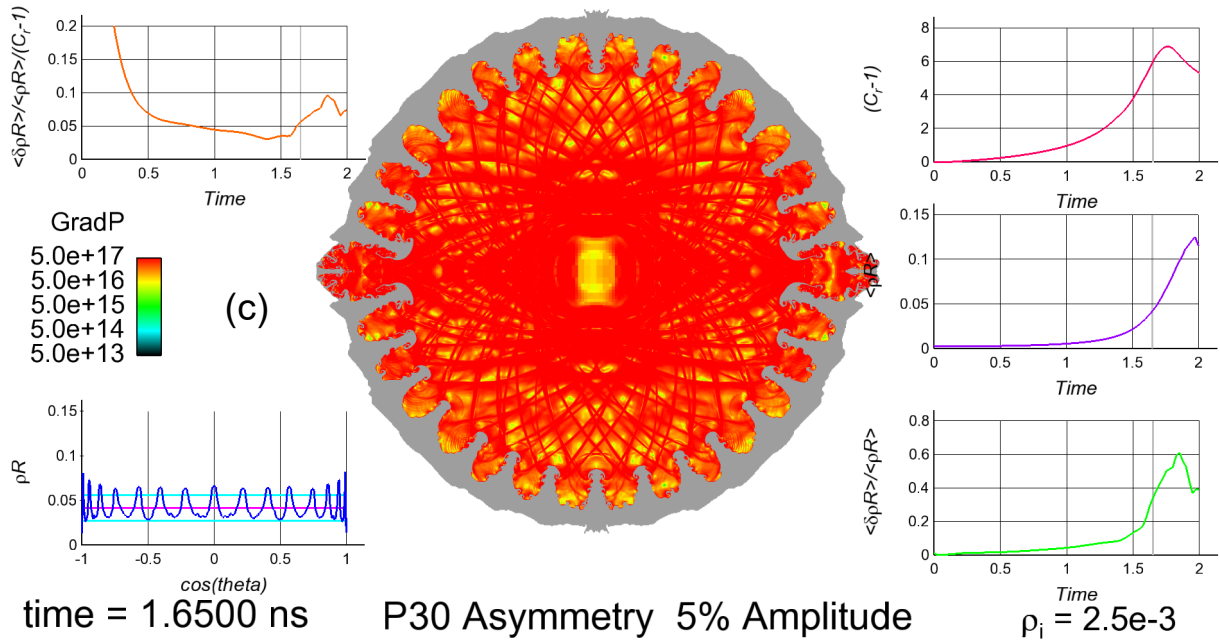

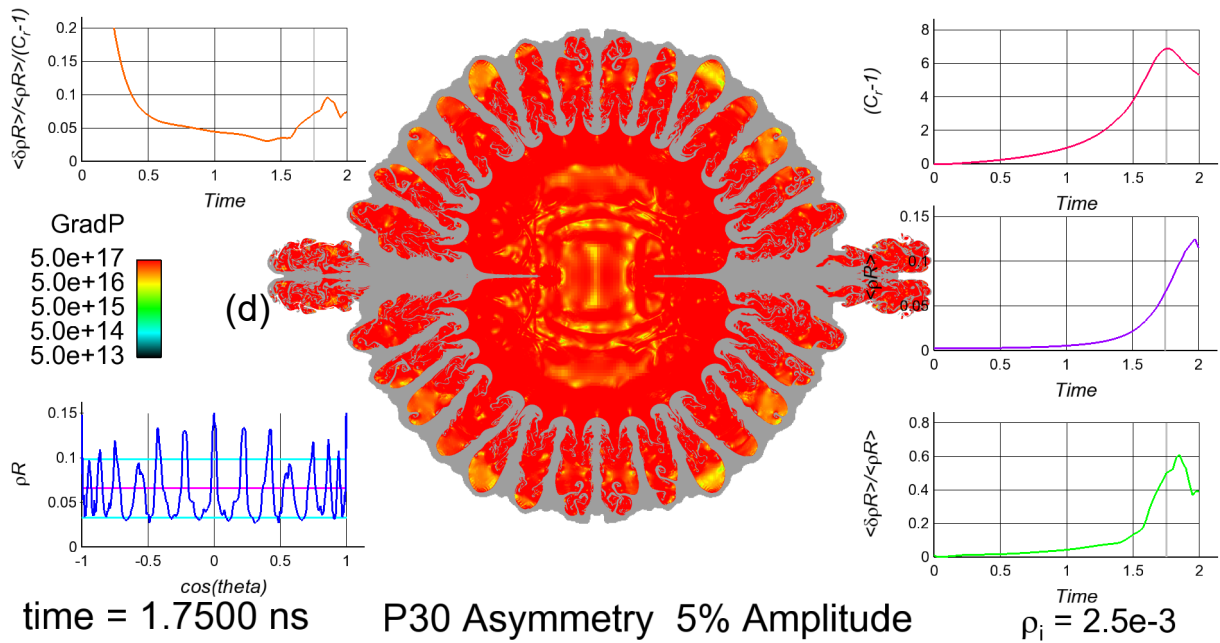

**Fig. 10. Four time snapshots from a 2D RAGE simulation of the P30 capsule with $a_{30}$ = 0.05 illustrating the rapid growth in $\rho R$ asymmetries of the CH shell.**

Fig. 10(a) shows a simulation time $t$ = 1.4 ns slightly after the 1st collision of the reflected gas shock with the incoming CH shell. The angular distribution of ρR in the lower left plotter shows only low amplitude growth of the ρR asymmetry in the period prior to the 1st collision of the reflected gas shock. The time history plot of <δρR>/<ρR> in the lower right plotter shows a sharp increase in the slope of the <δρR>/<ρR> growth curve at the time of the 1st reflected shock collision with the CH shell. <δρR>/<ρR> = 0.086 at this time.

Fig. 10(b) shows a simulation time $t$ = 1.575 ns slightly before the 2nd collision of the reflected gas shock with the incoming CH shell. The angular distribution of ρR in the lower left plotter shows noticeable growth as a result of the 1st shock collision. The time history plot of <δρR>/<ρR> in the lower right plotter shows a much sharper increase in the slope of the <δρR>/<ρR> growth curve at the time of the 2nd shock collision. <δρR>/<ρR> = 0.172 at this time.

Fig. 10(c) shows a simulation time $t$ = 1.65 ns near the time of the 3nd shock collision although by this time the shock reflections in the 2D simulation are so complex that a well defined main gas shock is no longer clearly evident. Here the angular distribution of ρR in the lower left plotter shows substantial growth in the ρR asymmetry as a result of the 2st shock collision. <δρR>/<ρR> = 0.339 at this time.

Finally, Fig. 10(d) shows a snapshot at stagnation, $t$ = 1.75 ns. The angular distribution of ρR in the lower left plotter shows that the ρR asymmetry has grown to large amplitude with <δρR>/<ρR> = 0.500 at this time.

The sequence of snapshots in Fig. 10 illustrate how repeated collisions of the reflected gas shock with the incoming shell amplify the ρR asymmetries until by stagnation fractional rms asymmetries of 50% are observed in the CH shell. Thus, a 5% asymmetry in the drive is seen to produce a 50% asymmetry in the ρR of the shell at stagnation for a convergence ratio of only 8.

In their proton spectrometry measurements of ρR asymmetries in OMEGA experiments, Li et al.[2] have attempted to correlate the measured fractional rms asymmetries <δρR>/<ρR> observed in the CH shell with fractional rms asymmetries <δI>/<I> in the direct drive laser illumination. The relationship they found is given by,

$$\frac{\langle \delta \rho R \rangle}{\langle \rho R \rangle} \approx \frac{1}{2}(C_r - 1)\frac{\langle \delta I \rangle}{\langle I \rangle} \tag{4}$$

This relationship is based upon a simple 1D scaling argument for the growth of ρR asymmetries from initial velocity perturbations in the shell. It suggests that the scaled fractional asymmetry

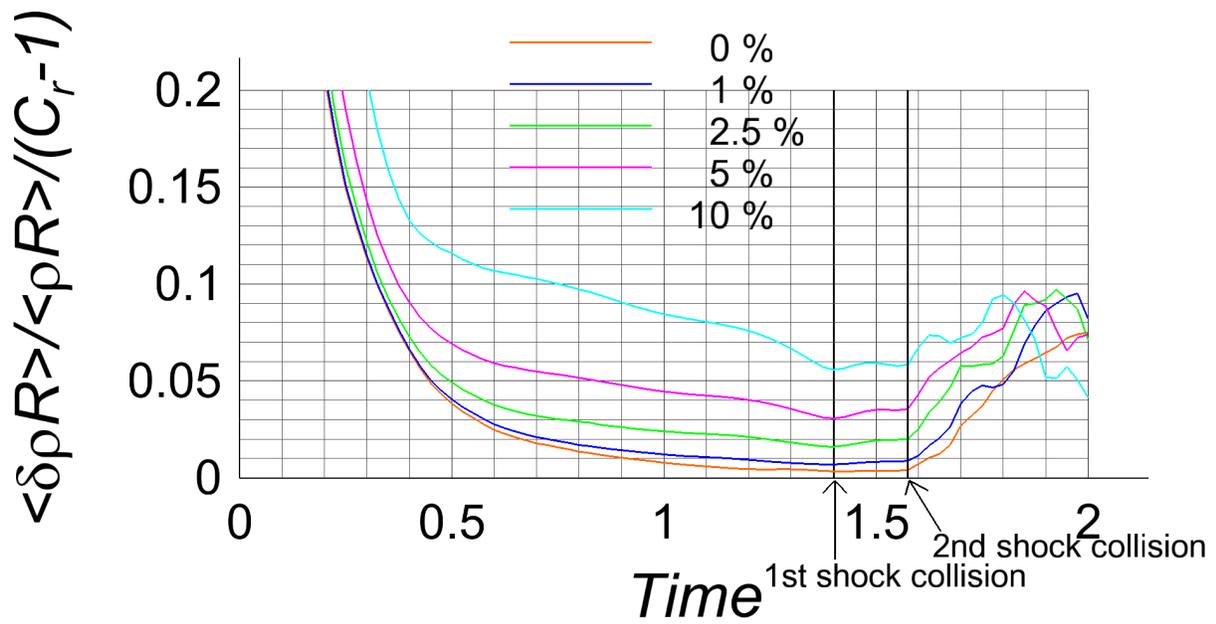

Fig. 11. Time history of the scaled fractional $\rho R$ asymmetry of the CH shell in 2D RAGE simulations of the P30 capsule with $a_{30}$ values of 0.00 (spherically symmetric), 0.01, 0.025, 0.05 and 0.10.

<δρR>/<ρR>/($C_r$ – 1) should be of order one half the amplitude of the drive asymmetry. Fig. 11 shows a time history plot of this quantity for five different 2D RAGE simulations with imposed asymmetries of 0, 1, 2.5, 5 and 10% amplitude. The time of the 1st reflected gas shock collision with the incoming shell is indicated in the figure. It can be readily seen from Fig. 11 that the scaled fractional asymmetry in ρR is indeed approximately equal to one half the amplitude of the imposed drive asymmetry at the time of this 1st shock collision. Note, however, that effects of the multiple decelerating shocks destroy this simple scaling relationship and lead to fractional asymmetries at stagnation that are much larger than would be expected from the simple 1D scaling argument.

### 2.1.3 Effect of Varying the Initial Gas Fill Density and Increasing Convergence Ratio

In this section we consider the effect of varying the initial gas fill density on the behavior of the implosion. To study this effect we performed eight different 2D RAGE simulations of the implosion of our idealized OMEGA capsule with a 5% imposed drive asymmetry but with initial gas fill densities ranging from $\rho_i = 2.5 \times 10^{-2}$ g/cm$^3$, 10 times the nominal fill density, down to $\rho_i = 5.0 \times 10^{-4}$ g/cm$^3$, one fifth the nominal fill density. Reducing the initial gas fill has the effect of increasing the final convergence ratio achieved at stagnation. However, this convergence ratio cannot be increased indefinitely because at some point in the process the capsule is completely disrupted and the integrity of the DT gas is total compromised.

Figs. 12(a) though 12(h) show snapshots of the eight different 2D RAGE simulations with decreasing gas fill at the time of minimum gas volume in each of the simulations. The gas at minimum volume has been divided into two regions, a perturbed region that is heavily penetrated by fingers of shell material, and a relatively unperturbed interior core of gas. This division is approximate, and is intended only to provide an estimate of the fraction of total gas volume which is perturbed and unperturbed in each case. The percentage of the total gas volume that is unperturbed and perturbed is indicated for each simulation along with the approximate value of ($C_r$ – 1) computed by taking $C_r = R(t)/R_0$ where the effective interface radius in 2D is computed from the DT gas volume using $R(t) = (3 \cdot V_{gas} / 4\pi)^{1/3}$ in the usual manner. The radius of this unperturbed gas region is also indicted for each snapshot. Note that the same magnification is utilized in each of these snapshot so that the snapshots can be compared directly with one another.

Fig. 12(a) shows the case with 10 times nominal fill density. In this case a convergence ratio of 3.7 is achieved at stagnation and only 21% of the total gas volume is perturbed by the intrusion of fingers of shell material. As the fill density is reduced the percentage of the total gas volume that is perturbed by fingers of shell material rises rapidly. Fig. 12(e) shows the case with the nominal fill density of $\rho_i = 2.5 \times 10^{-3}$ g/cm$^3$ where 70% of the total gas volume is perturbed and the convergence ratio is 7.9 . Fig. 12(g) shows the simulation with 2/5 the nominal gas fill. Here the percentage of perturbed gas volume has risen to 94% and the convergence ratio is 12.6 . Beyond this point it is no longer possible to meaningfully define an unperturbed core of gas as the final snapshot of Fig. 12(h) demonstrates. In the cases considered here it is not possible to exceed convergence ratios of 12 to 13 because the capsule is completely disrupted and the integrity of the gas is lost. It is interesting to note that in the experiments of Li et al.[2] no convergence ratios above 12 to 13 were experimentally achieved despite substantial reductions to the fill pressure of the CH capsules.

Fig. 12 also illustrates the computational challenges of accurately simulating very high convergence flows with RAGE. The snapshots of Figs. 12(f) through 12(h) show that for convergence ratios of 11.3 and above, unphysical flattening of the capsule along the 45º directions becomes increasingly pronounced. Accurate simulation of ICF implosions with convergence ratios in the range of 30 to 40 relevant for ignition targets without imposing an artificially high degree of implosion symmetry remains to date an unsolved problem of ICF simulation.

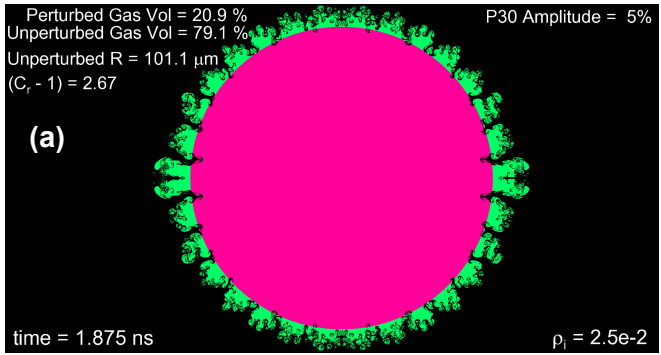
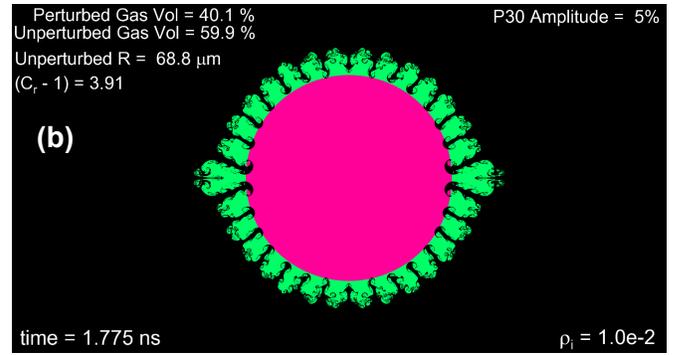
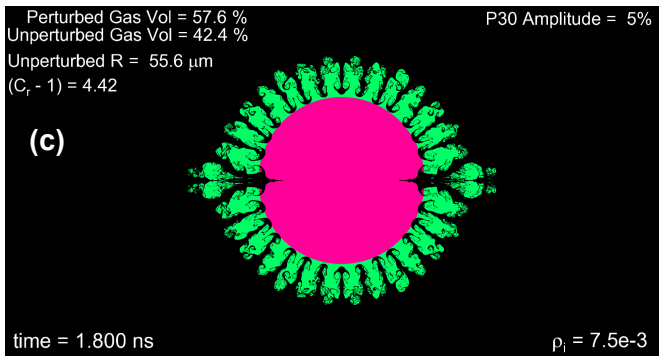
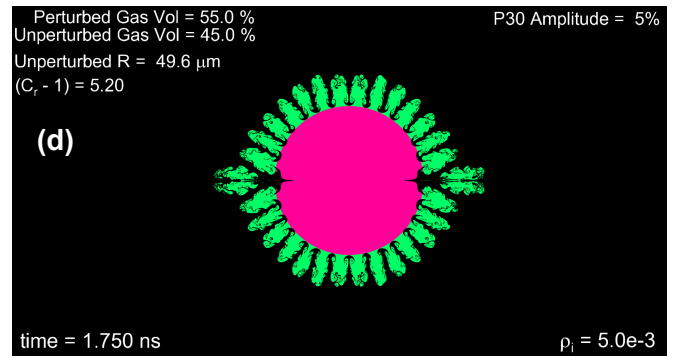
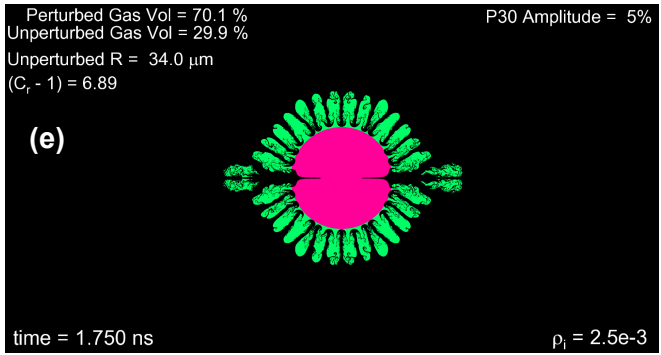
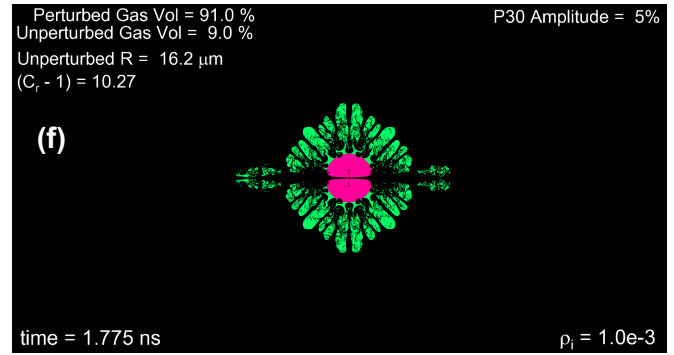
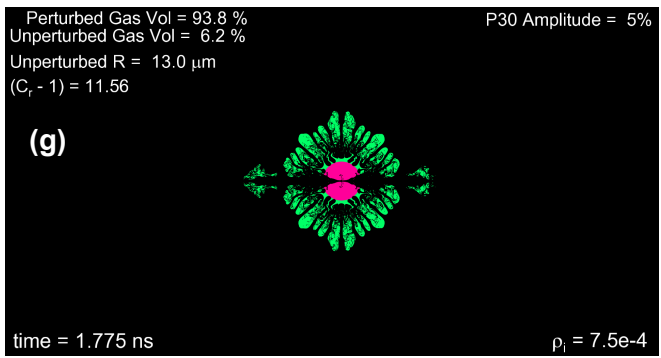
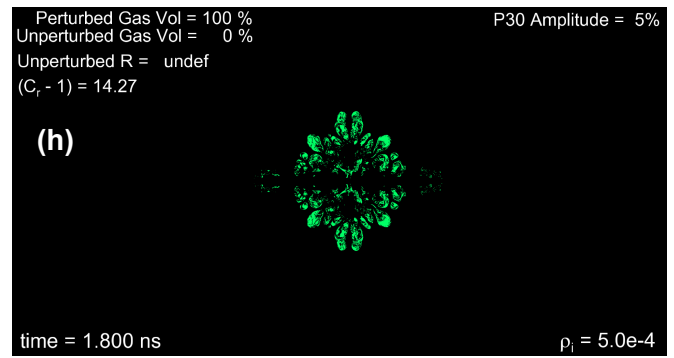

**Fig. 12. Snapshots at the time of minimum volume from eight different 2D RAGE simulations of the P30 capsule with $a_{30}$ = 0.05 but with different values for the initial gas fill density $\rho_i$ illustrating the disruption of the capsule for convergence ratios above roughly 13.**

### 2.1.4 Summary of 2D RAGE Simulation Results for the Idealized OMEGA Capsule

We can briefly summarize the results of our 2D RAGE simulations of the idealized OMEGA capsule implosion as follow. For a spherically symmetric drive, the implosion remains to a high degree spherical until stagnation. At stagnation the CH/gas interface exhibits the Rayleigh-Taylor structure one should expect when the system becomes physically RT unstable. For all amplitudes of asymmetric drive from 1% to 50% the density perturbations in the CH shell created by the drive asymmetry are amplified by Bell-Plesset related convergence effects. When acted upon by the outgoing reflected gas shock these perturbations act as seeds for the formation of Richtmyer-Meshkov fingers of shell material which penetrate the DT gas. The formation of these fingers is an essentially deterministic process with the number and angular position of the fingers being determined by the asymmetry of the drive. Even for 1% asymmetry and the convergence ratio of 8 achieved in these simulations, a clear pattern of 15 fingers of shell material are observed in the gas at stagnation. This process is very similar to the one described by Li *et al.*[2] in connection with the ρR asymmetries of imploded shells observed in their charged-particle spectrometry experiments on OMEGA.

The 2D simulations also show that the same drive asymmetry which leads to the formation of the Richtmyer-Meshkov fingers also results in the formation of coherent vortical structures in the DT gas. Strong, counter-rotating vortex rings are observed to form in close proximity both in the body of the DT gas and in the growing bubbles of DT gas which form between the Richtmyer-Meshkov fingers. It is well known that such vortical structures are subject to the unstable growth of azimuthal waves in 3D. There are two general classes of such azimuthal instabilities of counter-rotating vortex rings. The first is the short wavelength Widnall[6] instability with wavelengths along the azimuthal direction of the rings that are of the order of twice the transverse size of a vortex core. The second is the long wavelength Crow[7] instability between a pair of counter-rotating vortex rings with a wavelength along the azimuthal direction that is typically 6 to 8 times the separation distance between the rings. Generalizations of these simple idealized configurations involving many counter-rotating vortex rings interacting simultaneously are also possible. In the next section we investigate the stability of these strong, coherent vortical structures directly by performing 3D simulations initialized from our 50% asymmetrically driven 2D RAGE simulation.

### 2.2 Linked 3D RAGE Simulations of the Late Time Implosion

In our initial presentation of late time 3D RAGE simulation results for our idealized OMEGA implosion, we will focus on the case in which the P30 drive asymmetry has the relatively large amplitude of 50%. We have chosen this large amplitude for the P30 component in order to exaggerate the effect of the asymmetry on the implosion and to more clearly exhibit the mechanism that links the asymmetry to the development of turbulent jets in the fuel. Later we will present results from a 3D RAGE simulation with a P30 drive asymmetry but with only a 5% imposed asymmetry. This later simulation demonstrates quite clearly that even with only a modest 5% asymmetry and a low convergence of 8, turbulent gas jets can still form in a time that is relevant to the "compression burn". We begin with a consideration of the detailed simulation results for the 50% case.

A central property of the implosion that we are not directly modeling is the actual Reynolds number of the flow. Our 3D RAGE simulations are Eulerian and as such the shortest resolved wavelengths are limited by the numerical dissipation on the AMR grid rather than a physical parameter such as the viscosity. The actual Reynolds number of the flow is likely much too large to permit a direct numerical simulation using the Navier-Stokes equations. In addition detailed modeling of the viscosity would require accurate calculation of the temperatures. The high temperatures in the center may indeed lead to high viscosity but electron conduction in the small regions between the pusher fingers would likely lead to lower temperatures and much lower viscosity making the bubble regions late in time high Reynolds number flow. A qualitative understanding of at least some of the possible aspects of high Reynolds number flow in this system can be obtained by performing a spatial resolution study using 3D RAGE. The RAGE code allows the user to directly control the maximum level of mesh refinement in regions adjacent to material interfaces. This capability is particularly useful in the current problem since much of the vorticity generation and evolution observed in this problem occurs in the spatial regions near the CH/gas interface as a result of the interaction between the reflected gas shocks and the interface. Using 3D RAGE we performed a series of three simulations with increasing AMR spatial resolution near the CH/gas interface, all of which were initialized from the same 2D RAGE implosion at $t = 1.4$ ns. The first of these 3D RAGE simulations was linked from the 2D RAGE simulation of Figs. 2 and 3 at $t = 1.4$ ns, spun into 3D as an octant of the full problem and continued forward with a fixed maximum AMR spatial resolution near the CH/gas interface of 0.20 μm from link time out to a final problem time of $t = 1.938$ ns, a time well past minimum volume. In the second 3D RAGE simulation, the calculation was again linked at $t = 1.4$ ns and run from link time to $t = 1.5$ ns with a maximum AMR resolution of 0.20 μm. Then at $t = 1.5$ ns the maximum AMR resolution was increased to 0.1 μm and the problem was continued out to a final problem time of $t = 1.75$ ns, corresponding to the time of stagnation in the 1D RAGE simulation. In the third 3D RAGE simulation, the problem was again linked at $t = 1.4$ ns and run from t = 1.4 ns to t = 1.5 ns with a maximum AMR resolution of 0.20 μm. From $t = 1.5$ ns to $t = 1.6$ ns the maximum AMR resolution was increased to 0.10 μm. Finally, at $t = 1.6$ ns the AMR resolution was further increased to 0.05 μm and the problem was continued out to a final problem time of $t = 1.71$ ns, corresponding to a time just prior to the time of stagnation in the 1D RAGE simulation. By $t = 1.71$ ns, the final problem time achieved, the total number of AMR grid cells in 3D had grown to nearly $1 \times 10^9$. For convenience we refer to these three different 3D RAGE simulations as the 0.20 μm, 0.10 μm and 0.05 μm simulations respectively.

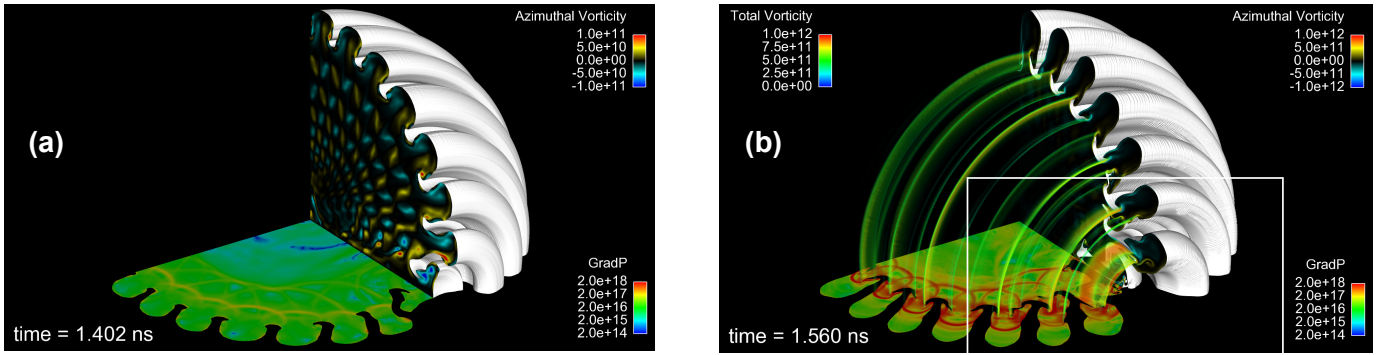

**Fig. 13.** Snapshots at (a) link time $t = 1.40$ ns and (b) at a later time t = 1.56 ns, the time of the 2nd collision of the reflected gas shock with the incoming CH capsule surface, for the linked 0.05 μm 3D RAGE simulation of the idealized OMEGA capsule with $a_{30} = 0.50$. In panel (b) the reflected gas shock can be seen traveling up the developing gas bubbles depositing sheets of vorticity which are visible in the volume-rendered representation of the total vorticity.

The procedure followed in all three of our 3D RAGE simulations was to run our P30 implosion simulation of the OMEGA capsule from $t = 0$ out to a link time of $t = 1.4$ ns as an axisymmetric 2D RAGE simulation. Then at the chosen link time of $t = 1.4$ ns we rotated one quadrant of the 2D axisymmetric problem into 3D to create a 3D octant version of the axisymmetric data. Figure 13(a) shows a 3D view of the octant at link time. This octant was used to initialize a 3D RAGE problem to continue the simulation to late time. An octant of the full capsule geometry was chosen to maximize the spatial resolution achieved in the 3D simulations.

The chosen link time of $t = 1.4$ ns corresponds to a time just after the first collision of the outgoing reflected gas shock with the incoming CH/gas interface, a time just after the beginning of significant enstrophy production in the 2D 50% asymmetry simulation as Fig. 8 illustrates. And since we needed to maintain high spatial resolution in the DT gas, linking at a significantly earlier time would have required running a much larger 3D simulation since the DT gas volume decreases rapidly during the implosion. Thus, the choice of $t = 1.4$ ns for the link time represents a balance between physics fidelity and computational practicality for the 3D problem.

Figure 13(b) shows a view of the capsule octant at a slightly later time $t = 1.56$ ns, the time of the 2$^{nd}$ collision of the reflected gas shock with the incoming CH shell, in the 0.05 μm 3D RAGE simulation of capsule implosion. In Fig. 13(b) only the central gas region of the 3D octant is shown with the surrounding CH shell material removed. The white surface represents the CH/gas interface. The vertical face of the gas is colored by the azimuthal vorticity and the horizontal face is colored by the gradient of pressure. The total vorticity in the interior of the gas is displayed in volume-rendered representation with a transfer function for color and opacity chosen to visualize regions of the flow with total vorticity above $5 \times 10^{11} \sec^{-1}$, a technique which is effective in revealing the dynamical evolution of the vortex cores in the interior of the gas bubbles. In Fig. 13(b) the reflected gas shock can be seen traveling up the gas bubbles between the developing fingers of shell material depositing sheets of vorticity of opposite sign along the opposite sides of the bubbles that are visible in the volume-rendered representation of the total vorticity.

Fig. 14 focuses in on the spatial region of Fig. 13(b) near the polar axis of the capsule enclosed by the white rectangle. Figs. 14(a)–(h) are a sequence of eight time snapshots from the 0.05 μm 3D RAGE simulation that show a close up view of the late time evolution of the vorticity in the gas bubbles nearest the polar axis of the capsule illustrating how the growth of azimuthal instabilities on the strong, counter-rotating vortex rings present in the gas bubbles leads to turbulence in the bubbles and the resulting interpenetration of CH and gas.

In Fig. 14(a) vortex sheets of opposite sign can be seen along opposite sides of the gas bubble nearest the polar axis. These sheets are pushed together by the radial convergence of RM fingers and are quickly rolled up into counter-rotating vortex rings that are trapped in the bubble and that immediately begin to undergo azimuthal instability growth in 3D (Figs. 14(b)–(c)) with a long wavelength Crow instability[7] clearly evident. This instability first appears in the interior of the gas and it is only when the increasingly turbulent vorticity of the gas flow impacts the CH surface (Figs. 14(e)–(h)) that fully 3D interpenetration and mixing occurs. This process may be envisioned as the injection of a gas jet into the bubble though the constriction caused by convergence of the RM fingers and the impact of this turbulent jet on the CH shell wall. A similar process can be seen occurring in the other bubbles as well with azimuthal instabilities of the short wavelength Widnall[6], long wavelength Crow[7] and combined type[8] all contributing to the very rapid evolution of the coherent vortical structures trapped in the bubbles to a fully turbulent state. As time progresses in the sequence of Fig. 14 the vortical structures in the bubble become increasingly complex and vortical structures of smaller and smaller spatial scale appear. In animations of these vortical structures we observe reconnection of vortex loops and it appears that repeated reconnection of closed vortex loops in 3D is the process which drives the cascade to smaller and smaller length scales observed in the panels of Fig. 14.

Our linked P30 simulation is not intended to model the results of an actual OMEGA experiment but rather to demonstrate the generic effects of an asymmetry in the pressure drive in a simple capsule geometry of physical interest. For this purpose we have deliberately chosen to impose a perturbation from spherical symmetry in the pressure drive which is simple, axisymmetric and relatively large and whose symmetry is constant in time. As we have seen, this initial drive asymmetry induces non-radial motions of the gas leading to the formation of vortex sheets in the gas. These vortex sheets are then rolled up into counter-rotating vortex rings by the outgoing reflected gas shock. The reflected gas shock also deposits intense sheets of azimuthal vorticity of opposite sign on nearby regions of the CH/gas interface. These vortical structures are unstable to azimuthal instabilities in 3D and quickly evolve to a fully turbulent state in which the turbulent vorticity in the gas adjacent to the CH/gas interface drives the interpenetration of gas and shell material. This same drive asymmetry also imprints a perturbation on the plastic shell by non-radial flow away from high pressure regions of the asymmetric drive into regions with low drive pressure. These density enhancements then grow due to convergence (the Bell-Plesset effect). Later in time, depending upon the degree of convergence and the amplitude of the initial asymmetry, these body perturbations induce surface perturbations in the CH/gas interface. These perturbations act as seeds for the Richtmyer-Meshkov instability as the outgoing reflected gas shock collides with the interface. It is the radially converging growth of these Richtmyer-Meshkov fingers that ultimately lead to the formation of the final shape of the gas region. Thus, large asymmetries in the pressure drive produce both the Richtmyer-Meshkov fingers which determine the shape of the gas region and the vortical structures in the gas which lead to the gas turbulence and mix via 3D azimuthal instabilities.

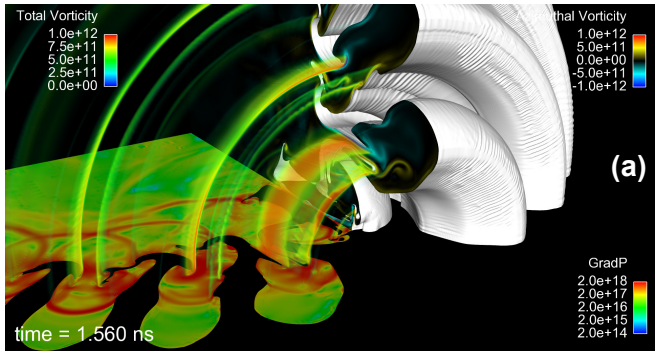 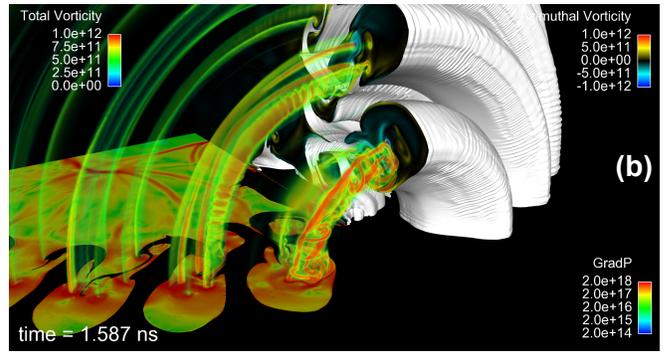
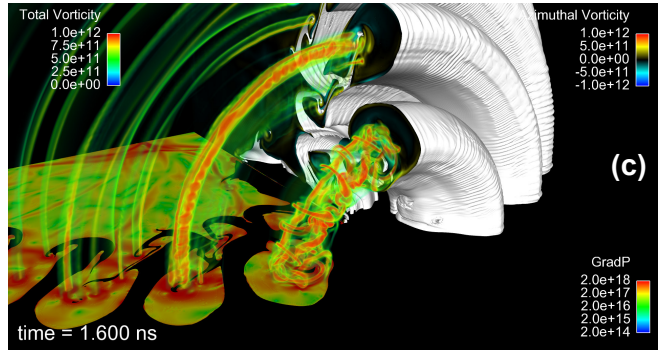 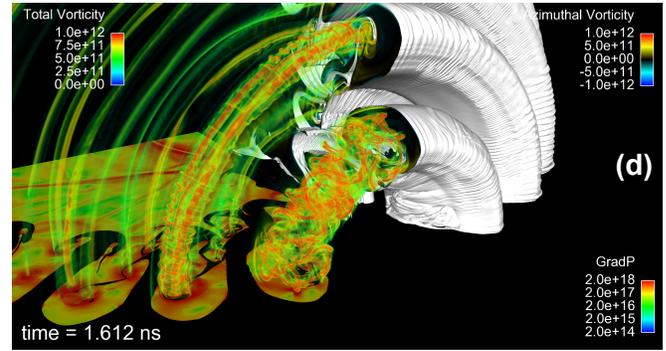
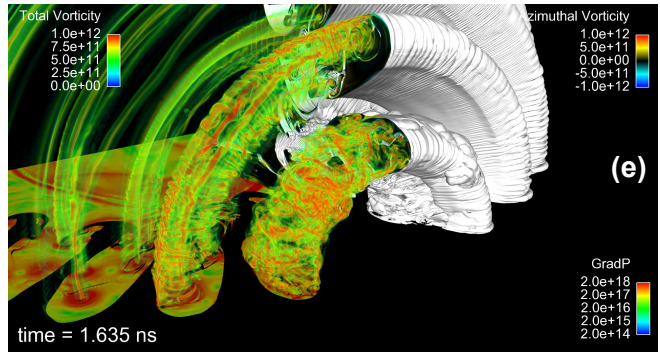 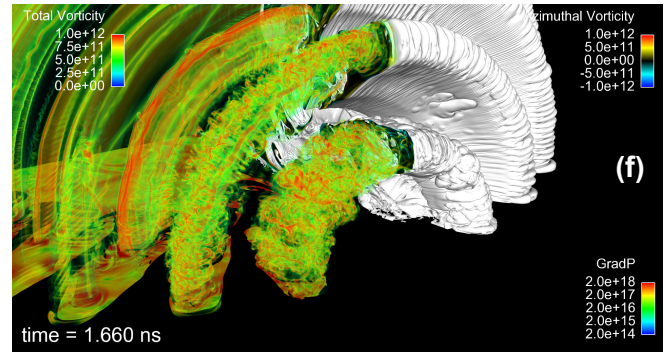
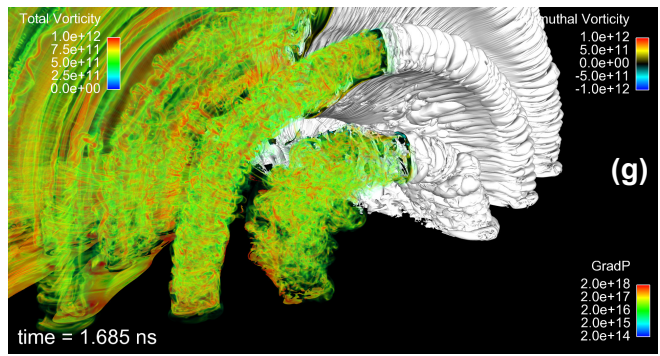 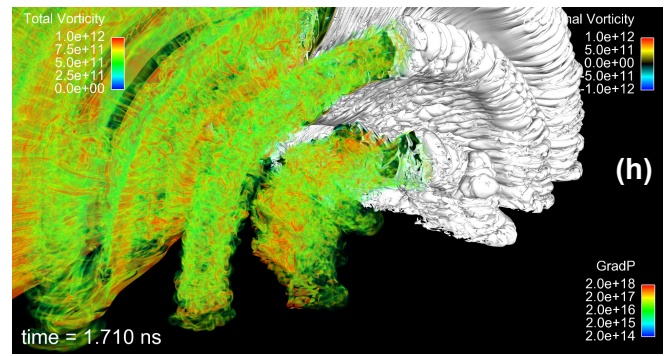

**Fig. 14. Eight time snapshots from the 0.05 μm 3D RAGE simulation of the P30 OMEGA capsule. Close up view of the development of the turbulence in the two bubbles nearest the polar axis.**

In this simple example we can see, clearly exhibited, a mechanism connecting the asymmetry of the pressure drive on the capsule with the development of turbulence and mix. This mechanism is fully three dimensional in nature. The asymmetry of the drive leads to the formation of a well-defined pattern of coherent vortical structures, counter-rotating sheets and rings, in the gas. These structures are unstable to azimuthal instabilities in three dimensions and it is the unstable evolution in 3D of these coherent structures that lead to the development of the gas turbulence and mix.

Further, it can be seen in this simple example that both the number and angular distribution of the fingers and the spatial distribution of the gas turbulence and its associated mix are largely deterministic in nature and are a direct result of the deviations from spherical symmetry in the pressure drive. The unstable growth of the gas turbulence from coherent vortical structures is a stochastic process. But the nature and spatial distribution of these coherent vortical structures is determined by the asymmetry of the pressure drive.

### 2.2.1 Spatial Resolution Study of 3D RAGE Simulations

An overview of the results from a resolution study comparing the 0.20 μm, 0.10 μm and 0.05 μm 3D simulations is summarized in Figs. 15(a) through 15(c) which show the same close up view of the bubbles nearest the polar axis for each of the three simulations at a fixed simulation time of $t$ = 1.71 ns, the last simulation time where data from all three simulation resolutions is available, as the spatial resolution of the 3D simulation is increased. In each snapshot the total vorticity in the interior of the gas is displayed in a volume-rendered representation with the same transfer function for color and opacity chosen to visualize regions of the flow with total vorticity above $5 \times 10^{11} \sec^{-1}$. The peak vorticity observed at $t$ = 1.71 ns in the simulations grows from $9.2 \times 10^{11} \sec^{-1}$ to $5.3 \times 10^{12} \sec^{-1}$ as the resolution is increased from 0.20 μm to 0.05 μm. As the resolution progresses from 0.20 μm in Fig. 15(a) to 0.05 μm in Fig. 15(c), the degree of turbulent development observed in the gas bubbles nearest the polar axis at a fixed simulation time steadily increases with a corresponding increase in the three dimensional structure of the CH/gas interface as a result of the interaction between the gas turbulence and the adjacent CH/gas interface consistent with the higher resolution calculations representing a higher effective Reynolds number. To the extent that the real Reynolds number is significantly higher than that of the calculation, then the expected level of three dimensional effects would also be elevated compared to these calculations.

A standard method for demonstrating that turbulence has been observed in a fluid simulation is to compute the power spectrum of the kinetic energy for the simulation and look for the appearance of an inertial subrange in the spectrum. An inertial subrange is a region of the spectrum in $k$ space over which energy cascades from longer to shorter wavelengths before being dissipated as heat either by viscosity in the case of a real fluid or by the effective dissipation terms of the numerical method as in the case of our current inviscid simulations. The kinetic energy spectrum in the inertial subrange typically has a power law dependence of the form $k^{-\alpha}$ where for homogeneous, isotropic turbulence this power takes the Kolmogorov value of $\alpha = 5/3$.

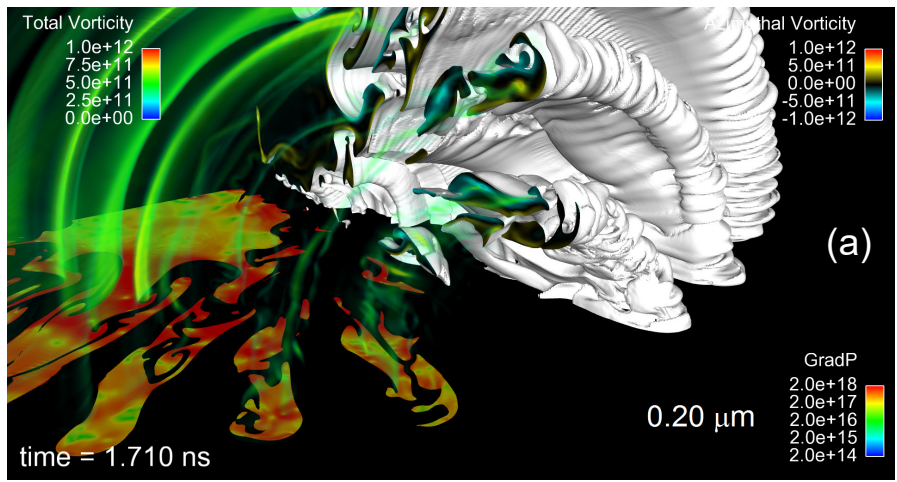
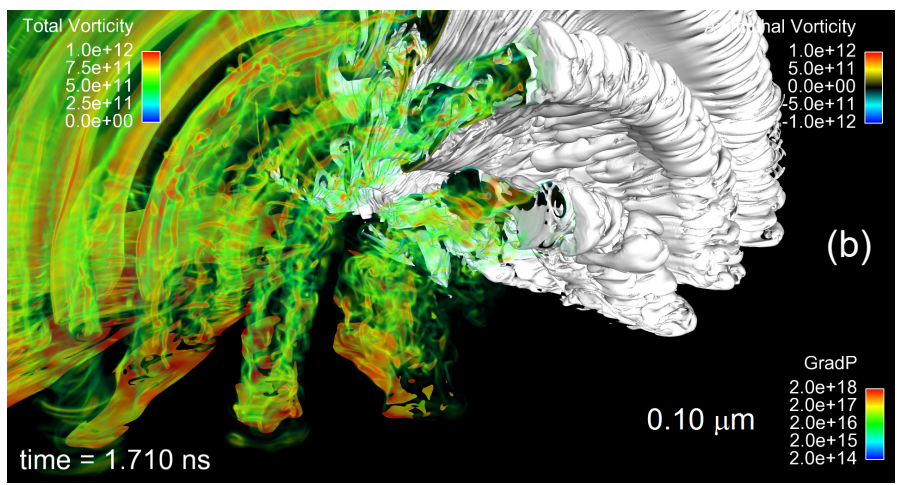
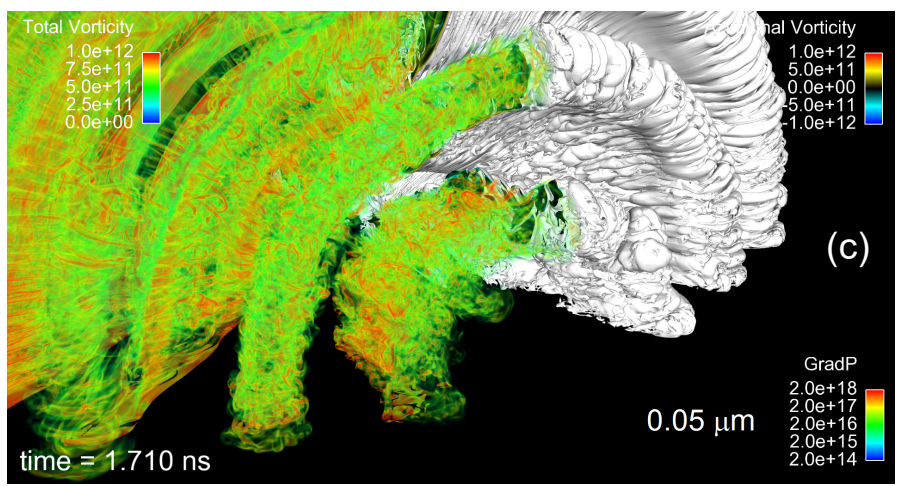

**Fig 15. Effect of spatial resolution on the development of turbulent structure at a fixed simulation time of *t* = 1.71 ns. The maximum permitted AMR resolution for each 3D Rage simulation is, from top to bottom, 0.20 μm, 0.10 μm and 0.05 μm.**

To demonstrate the presence of turbulence in our 3D RAGE simulation, our procedure was to select the fixed square spatial box 33 μm on a side shown in Fig. 16(a). This box was chosen to completely enclose the two turbulent bubbles nearest the polar axis. Values within the box for the density and the three Cartesian velocity components were re-sampled from the AMR mesh onto a uniform mesh whose spatial resolution was equal to the finest resolution used in the corresponding AMR mesh. The re-sampled uniform mesh data was then used to compute the power spectrum by the following method.

By Parseval's theorem the total kinetic energy within the box can be written in the form,

$$\int \tfrac{1}{2}\rho |\vec{u}|^2 d^3\vec{x} = \int \vec{H}(k)\cdot \vec{H}(k)^* d^3\vec{k} = \int E(k)dk \quad (5)$$

where $k = |\vec{k}|$

$\vec{H}(k)$ is the Fourier transform of the mass-weighted velocity $\sqrt{\dfrac{\rho}{2}}\vec{u}$ and

$$E(k) = 4\pi k^2 \vec{H}(k)\cdot \vec{H}(k)^* \quad (6)$$

is the spectral power of the kinetic energy between $k$ and $k + dk$.

Using the resampled data on the uniform mesh, an FFT was utilized to compute the discrete Fourier transform values for $\vec{H}(k)$ on a mesh in three dimensional $k$ space. $E(k)$ is then computed by averaging over the directions in $k$ space within a shell between $k$ and $k + dk$. Fig. 16(b) shows the resultant power spectra of the kinetic energy for each of the three simulations at spatial resolutions of 0.20 μm, 0.10 μm and 0.05 μm at time $t = 1.71$ ns.

The spectrum for the 0.20 μm simulation shows no clear evidence that an inertial subrange has yet developed in the turbulence growing in the bubbles near the polar axis for the 0.20 μm simulation by $t = 1.71$ ns. However, Fig. 16(b) shows that as the spatial resolution of the simulation is increased, an inertial subrange appears to emerge in the turbulence near the axis at $t = 1.71$ ns. In the spectrum for the 0.05 μm simulation at $t = 1.71$ ns, an inertial subrange is present with a Kolmogorov type $k^{-5/3}$ power law scaling from $k = 25000$ cm$^{-1}$ out to $k = 80000$ cm$^{-1}$ followed by a steep dissipation regime at higher values of $k$. The location of the start of the dissipation region at $k = 80000$ cm$^{-1}$ corresponds to a length scale of about 16 grid cells in the 0.05 μm RAGE simulation. The spectrum for the 0.10 μm resolution case shows a similar short inertial subrange which again starts at around $k = 25000$ cm$^{-1}$ and falls off into the dissipation regime at about $k = 40000$ cm$^{-1}$, again corresponding to a length scale of 16 grid cells in the 0.10 μm simulation. The 0.20 μm simulation at $t = 1.71$ ns cannot see this inertial subrange at all because the dissipation regime begins at $k = 20000$ cm$^{-1}$ in the 0.20 μm case. The results of this spectral analysis demonstrate that the turbulence near the polar axis observed at $t = 1.71$ ns in the 0.05 μm 3D RAGE simulation exhibits an inertial subrange with $k^{-5/3}$ power law scaling at a time prior to the stagnation time $t = 1.75$ ns obtained in the 1D RAGE simulation.

The appearance of turbulence with a Kolmogrov inertial subrange in our 0.05 μm 3D RAGE simulation is significant for ICF applications because it demonstrates how asymmetry can lead to hydrodynamic turbulence by way of instability of large coherent features in a time that is short enough to be of interest for degrading the compression burn of an ICF implosion. This result presents a plausible alternative picture to the usual paradigm that degradation in ICF implosions is caused by growth of small scale surface imperfections either by acceleration driven instabilities and/or shock induced instabilities. These results may also provide a plausible answer to the question of how the vortices at small scales in turbulent flows begin. One direct answer to this question is to note the appearance in our simulations of the small scale Widnall type waves visible in the volume-rendered representation of the total vorticity. These waves have wavelengths of the order of the vortex core size which is determined by the residual dissipation of the numerical simulation. Thus, from the large scale coherent structures significant wave activity can be generated near the dissipation region of the fluctuations by the appearance of short wavelength instabilities of the Widnall type. A different source of enhanced fluctuations at intermediate scale appears to result from longer wavelength Crow type instabilities of the large coherent structures. Such waves are seen, for example, in the $t = 1.587$ ns panel of Fig. 14(b). It has been proposed that these Crow instabilities lead to vortex reconnection loops with smaller scales than the original structures and this process then repeats as the energy in longer wavelengths cascades to the dissipation regime (Takakai and Hussain[10] and Kerr and Hussain[11]). We may, in fact, be observing this process in the 0.05 μm simulation discussed above. More work is required to clarify the role of repeated reconnection of vortex loops in creating the inertial cascade observed in our 3D ICF simulations. We do note, however, that both processes appear to be fast enough to populate the inertial range with a sizeable amount of fluctuations allowing the generation of the inertial range shown in Fig. 16(b). Both processes are the result of hydrodynamic instability of coherent large scale hydrodynamic structures created by asymmetry of the drive. Beautiful experimental observations and a comprehensive discussion of the development of both short wavelength and long wavelength azimuthal instabilities in a system of two counter-rotating fluid vortices may be found in the paper of Leweke and Williamson[8].

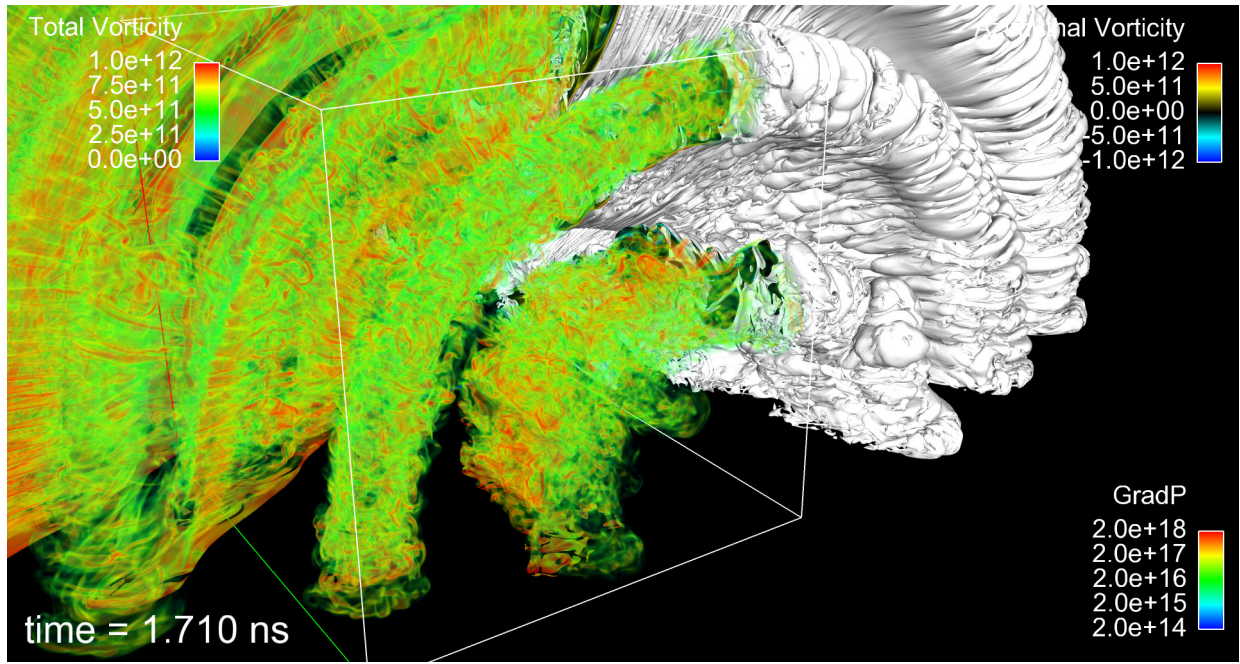

**Fig. 16(a). Box showing the sampling region selected for Fourier analysis. This box contains the two turbulent rings nearest the polar axis of the capsule.**

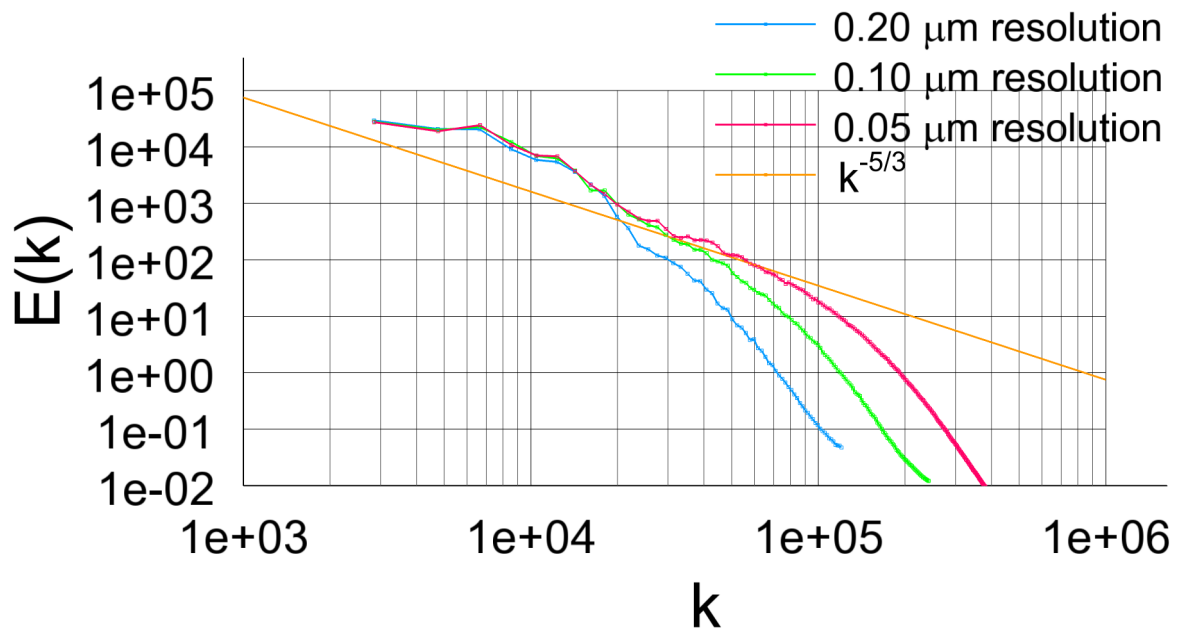

**Fig. 16(b). Power spectra of the total kinetic energy at *t* = 1.71 ns for the three spatial resolutions shown in Figs. 15(a) through (c). At a spatial resolution of 0.05 μm an inertial subrange is emerging that is consistent with a Kolmogorov-type spectrum.**

### 2.2.2 Enstrophy Production in the 3D RAGE Simulations

A more quantitative measure of the turbulent state of the gas in our 3D RAGE simulation is the total enstrophy defined as the volume integral,

$$\int \tfrac{1}{2} |\vec{\omega}|^2 d^3\vec{x} = \int \tfrac{1}{2} \{\omega_z^2 + \omega_r^2 + \omega_\phi^2\} d^3\vec{x} \quad (7)$$

where

$$\vec{\omega} = \nabla \times \vec{v}$$

is the vorticity of the velocity field $\vec{v}$ in the 3D simulation. ($\omega_z, \omega_r, \omega_\phi$) are the axial, radial and azimuthal components of vorticity in cylindrical coordinates $(z, r, \phi)$ and the integral is extended over the entire 3D simulation volume. Quantitative information on the evolution of the gas turbulence in the 3D RAGE simulations of Fig. 15 can be obtained by examining the time history of the total enstrophy as well as the separate contributions to the total enstrophy from the axial, radial and azimuthal components of the vorticity in Eqn. (7).

In Fig. 17(a) the total enstrophy and the azimuthal, radial and axial enstrophy are shown for the 3D RAGE simulation with fixed 0.20 μm maximum AMR resolution. For a fully axisymmetric simulation the total enstrophy would exactly equal the azimuthal enstrophy and the non-azimuthal radial and axial contributions to the total enstrophy would be identically zero. In Fig. 17(a) we see that the azimuthal enstrophy tracks the total enstrophy very closely until a time just slightly after $t = 1.6$ ns corresponding to the beginning of turbulence growth in the gas bubble nearest the polar axis. By $t = 1.71$ ns significant non-azimuthal radial and axial contributions to the total enstrophy have begun to be apparent and the time period from $t = 1.71$ ns to $t = 1.938$ ns corresponds to a period of rapid growth in the non-azimuthal components of the total enstrophy. During this same period the azimuthal component reaches a peak at $t = 1.77$ ns and then falls rapidly until by the end of the simulation at $t = 1.938$ ns all three components of the total enstrophy have become comparable in magnitude as would be the case in isotropic turbulence. For the maximum AMR resolution 0.20 μm simulation only a small amount of non-azimuthal enstrophy growth is observed by $t = 1.75$ ns, the time of stagnation in the 1D simulation of the implosion, consistent with Fig. 15(a) and 16(b).

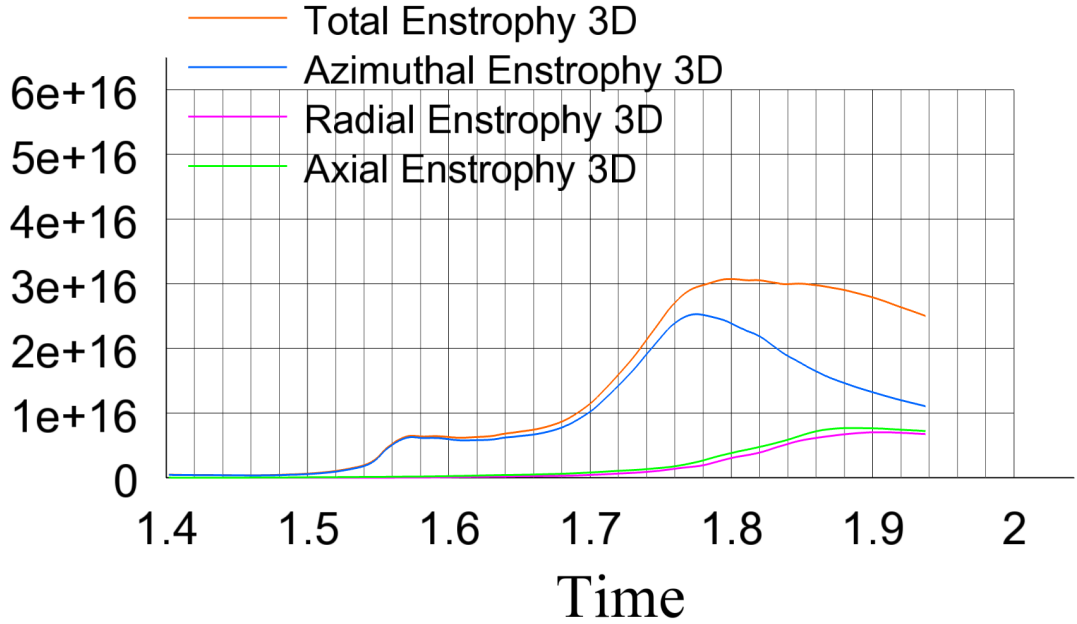

**Fig. 17(a). Time history for the total enstrophy and for all three components of the enstrophy of the gas for the 0.20 μm simulation in 3D. The axial and radial components track closely and grow as the turbulent rings develop. Enstrophy is in units of cm³/sec².**

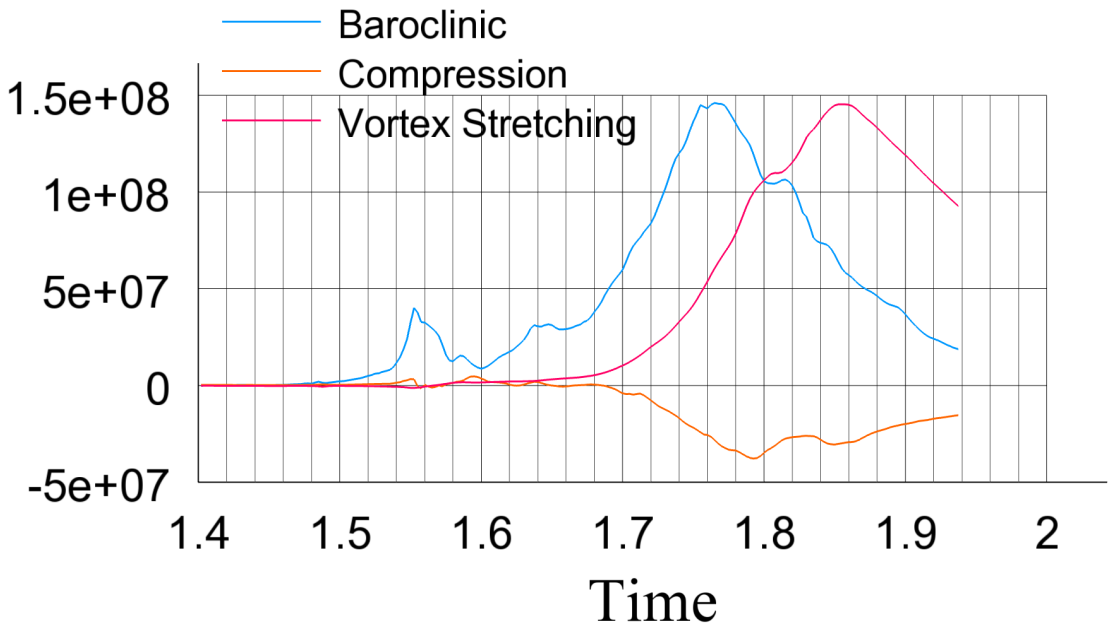

**Fig. 17(b). Time history for the three enstrophy source terms in the total enstrophy production Eqn. (9). The production of non-azimuthal enstrophy observed in Fig. 17(a) corresponds to the vortex stretching term.**

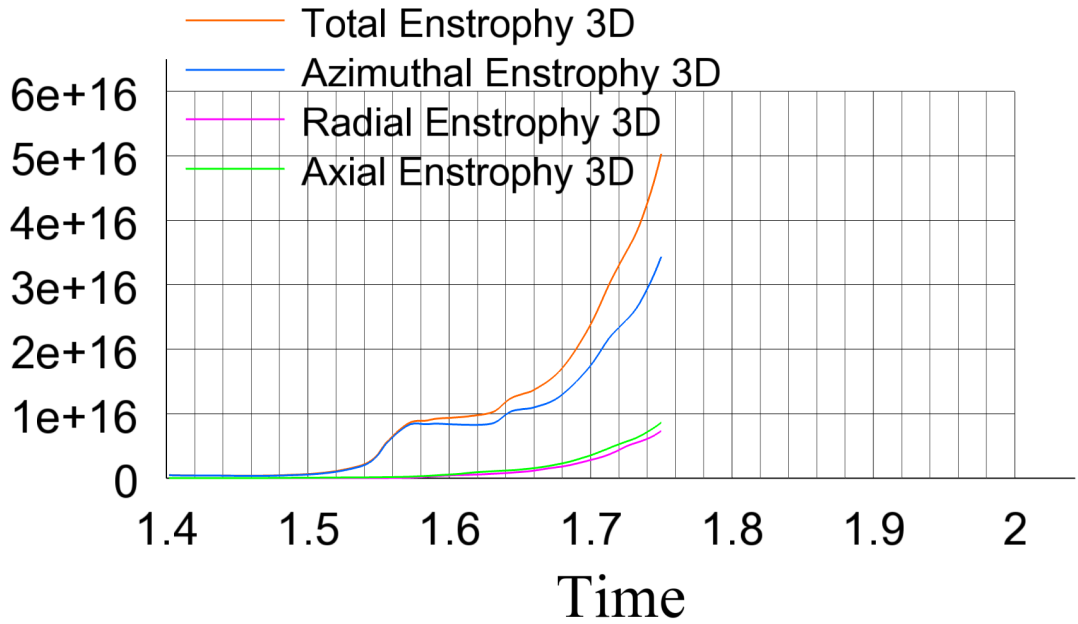

**Fig. 17(c).** Time history for the total enstrophy and for all three components of the enstrophy of the gas for the 0.10 $\mu$m simulation in 3D. Enstrophy is in units of cm$^3$/sec$^2$.

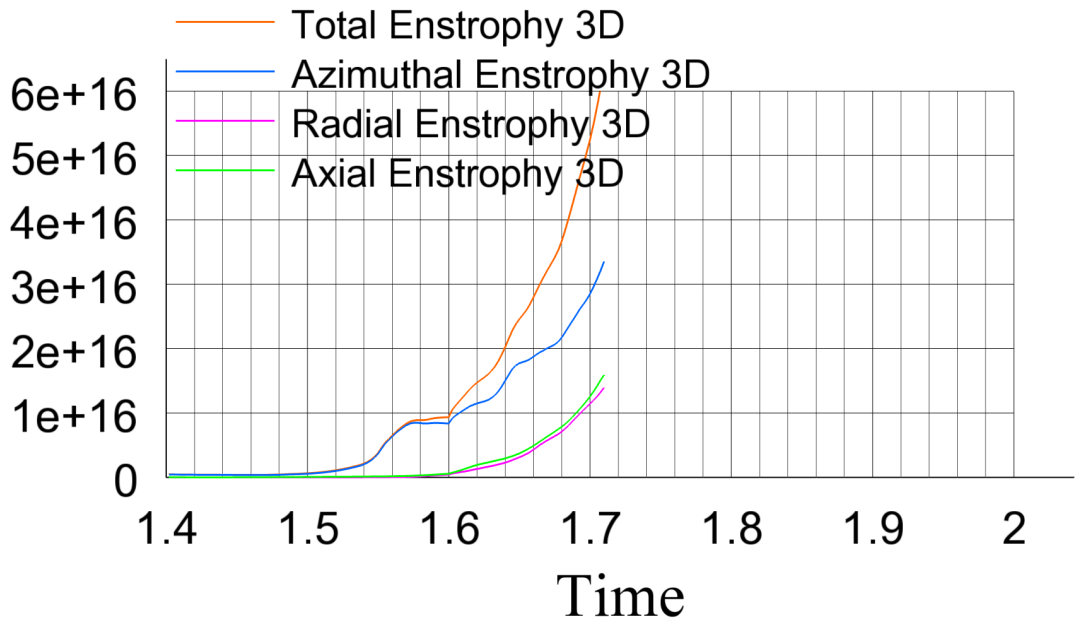

**Fig. 17(d).** Time history for the total enstrophy and for all three components of the enstrophy of the gas for the 0.05 $\mu$m simulation in 3D. Enstrophy is in units of cm$^3$/sec$^2$.

More detailed quantitative information can be obtained by considering the Euler equation for the enstrophy production. If we define the quantity $\Omega$ as,

$$\Omega \equiv \tfrac{1}{2}|\vec{\omega}|^2$$

Then the enstrophy production equation can written in the form,

$$\frac{\partial \Omega}{\partial t} + \nabla \cdot (\Omega \vec{v}) = \frac{1}{\rho^2} \vec{\omega} \cdot (\nabla \rho \times \nabla P) - \Omega(\nabla \cdot \vec{v}) + \omega^i S_{ij} \omega^j$$

(8)

where the quantity

$$S_{ij} = \frac{1}{2}\left(\frac{\partial v^i}{\partial x^j} + \frac{\partial v^j}{\partial x^i}\right)$$

is the strain rate tensor and summation over the repeated indices $i$ and $j$ is understood. No explicit dissipation terms have been included in Eqn. (8) which corresponds to the enstrophy production by the inviscid Euler equations. Such terms will, of course, arise from the numerical hydro method used in RAGE.

Integrating Eqn. (8) over the total volume of the simulation and applying the divergence theorem to the second term on the left to convert it to a surface integral over the boundary of the simulation volume yields a simple equation for the time rate of production of the total enstrophy,

$$\frac{\partial}{\partial t}\int \Omega\, d^3\vec{x} = \int \left[\frac{1}{\rho^2}\vec{\omega}\cdot(\nabla\rho\times\nabla P) - \Omega(\nabla\cdot\vec{v}) + \omega^i S_{ij}\omega^j\right] d^3\vec{x}$$

(9)

where the surface terms on the left side are identically zero because of the reflecting boundary conditions at the boundary of the simulation volume. The first term on the right is the baroclinic enstrophy production term, the second is the compression term and the third is the vortex stretching term. The vortex stretching term exists only in 3D and plays a dominant role in the development of three dimensional turbulence.

Fig. 17(b) shows the time histories for the baroclinic, compressible, and vortex stretching enstrophy production terms in Eqn. (9) obtained from our 3D RAGE simulation with fixed maximum AMR resolution of 0.20 μm. It can be seen that at early times the enstrophy production is predominantly baroclinic whereas at later time the increase in the radial and axial components of the enstrophy observed in Fig. 17(a) are associated with a significant enstrophy production contribution from the vortex stretching term. Significant growth of the vortex stretching contribution is seen beginning at roughly $t = 1.68$ ns and reaching a peak value at around $t = 1.855$ ns. Contributions from the compressible term remains relatively small at all times and are negative after minimum volume in the 3D simulation as would be expected.

A more quantitative characterization of the turbulence growth as a function of resolution can again be obtained by considering the time behavior of the non-azimuthal components of the enstrophy for each of the three resolutions considered. Recall that the non-azimuthal enstrophy components, the radial and axial enstrophy components, are initially zero at link time since each of our 3D simulations is initialized from an 2D axisymmetric simulation in which both the radial and axial components of the vorticity are identically zero. The non-azimuthal enstrophy components grow in time as a result of the development of fully three-dimensional turbulence and provide a useful quantitative measure of the degree of turbulence growth. It is therefore of interest to compare the time development of the non-azimuthal enstrophy components for each of the 3 resolutions.

The time history of total enstrophy and of all three enstrophy components for the 0.20 μm, 0.10 μm and 0.05 μm simulations are shown in Figs. 17(a), 17(c) and 17(d) respectively. Comparison of the time histories of non-azimuthal enstrophy components in Figs. 17(a), 17(c) and 17(d) show that the non-azimuthal enstrophy grows more rapidly in time as the spatial resolution of the simulation is increased. For example, comparing Figs. 17(a) and 17(c) shows that the total enstrophy at $t = 1.71$ ns in the 0.10 μm simulation is 2.1 times greater than the corresponding value in the 0.20 μm simulation, while the sum of non-azimuthal enstrophy components, radial plus axial, at $t = 1.71$ ns in the 0.10 μm simulation is 5.3 times greater than the corresponding value in the 0.20 μm simulation. Similarly, the total enstrophy at $t = 1.71$ ns in the 0.05 μm simulation is 2.22 times greater than the total enstrophy in the 0.10 μm simulation, while the sum of non-azimuthal enstrophy comonents at $t = 1.71$ ns in the 0.05 μm simulation is 3.83 times greater than the corresponding value in the 0.10 μm simulation at $t = 1.71$ ns. Overall, as the spatial resolution of the simulation is increased from 0.20 μm to 0.05 μm, the total enstrophy at $t = 1.71$ ns increases by a factor of 4.67, while the sum of the non-azimuthal enstrophy components at $t = 1.71$ ns increases by a factor of 20.37. The observed growth of the non-azimuthal enstrophy components further emphasizes the point that for higher resolution 3D RAGE simulations, the turbulence develops more rapidly in time.

### 2.2.3 3D RAGE Simulations of the 5% Amplitude Case

We now consider a 3D RAGE simulation for the case of a 5% imposed asymmetry in our simplified ICF implosion system. This simulation was performed using the same 2D-3D linking procedure previously utilized for the case of a 50% imposed asymmetry. Here the simulation was run with an AMR resolution of 0.05 μm near the CH/gas interface from the chosen link time of $t = 1.5$ ns. This later link time was chosen in order to focus on the late time hydrodynamics occurring near stagnation.

The sequence of eight time snapshots of the full capsule for the case of 5% imposed asymmetry is shown in Fig. 18. In each snapshot the white surface, as usual, represents the CH/gas interface. The vertical face of the gas is colored by the azimuthal vorticity and the horizontal face is colored by the gradient of pressure. The total vorticity in the interior of the gas is displayed in a volume-rendered representation with the same transfer function as that used in Fig. 14 for the 50% case.

Fig. 18(a) at $t = 1.675$ ns shows the implosion just after the 3$^{rd}$ shock collision with the CH/gas interface. Note the complex reflections off the perturbed CH/gas interface that can be seen in Fig. 18(a) traveling inward toward the center of the capsule. Part of the incident shock can also be seen still traveling in the radial direction toward the end of the gas bubbles in Fig. 18(a). Since the interface is strongly perturbed at this time, the incident shocks will in general be misaligned with the density gradients, producing significant shear layers at the CH/gas interface that are visible as the vortex sheets forming along the edges of the fingers in Fig. 18(a). As time progresses these sheet are, as we have already pointed out, rolled up into vortex rings of opposite sign in the azimuthal vorticity that are trapped together inside the developing gas bubbles.

Fig. 18(b) at $t = 1.700$ ns shows the collapse onto the capsule center of this complex reflected gas shock which bounces off the capsule center at about $t = 1.708$ ns.

Fig. 18(c) shows a slightly later time $t = 1.725$ ns with the resultant outgoing reflected gas shock again moving radially outward. Note by the time of Fig. 18(c) that both short and long wavelength unstable azimuthal waves have begun to grow on the vortex rings trapped in the two bubbles nearest the polar axis.

Fig. 18(d) at $t = 1.75$ ns is approximately at stagnation time and is also the time of the 4$^{th}$ shock collision of the reflected gas shock with the now highly perturbed CH/gas interface. The vortex rings trapped in the two gas bubbles nearest the polar axis have already become turbulent at this time and unstable waves have begun to grow to large amplitude in several other bubbles as well. The 4$^{th}$ gas shock collision can also be seen generating additional strong vortex sheets along the edges of the developing fingers.

The high pressure gas at the center of the capsule is driving high speed jets of gas through the now constricted bubble openings which can be viewed as "nozzles" forming turbulent jets of gas that traverse the developing gas bubbles. By $t = 1.768$ ns in Fig. 18(e) the turbulent jet in the bubble nearest the polar axis can be seen just beginning to collide with the CH/gas interface to produce as time progresses fully three dimensional interpenetration of the gas and CH.

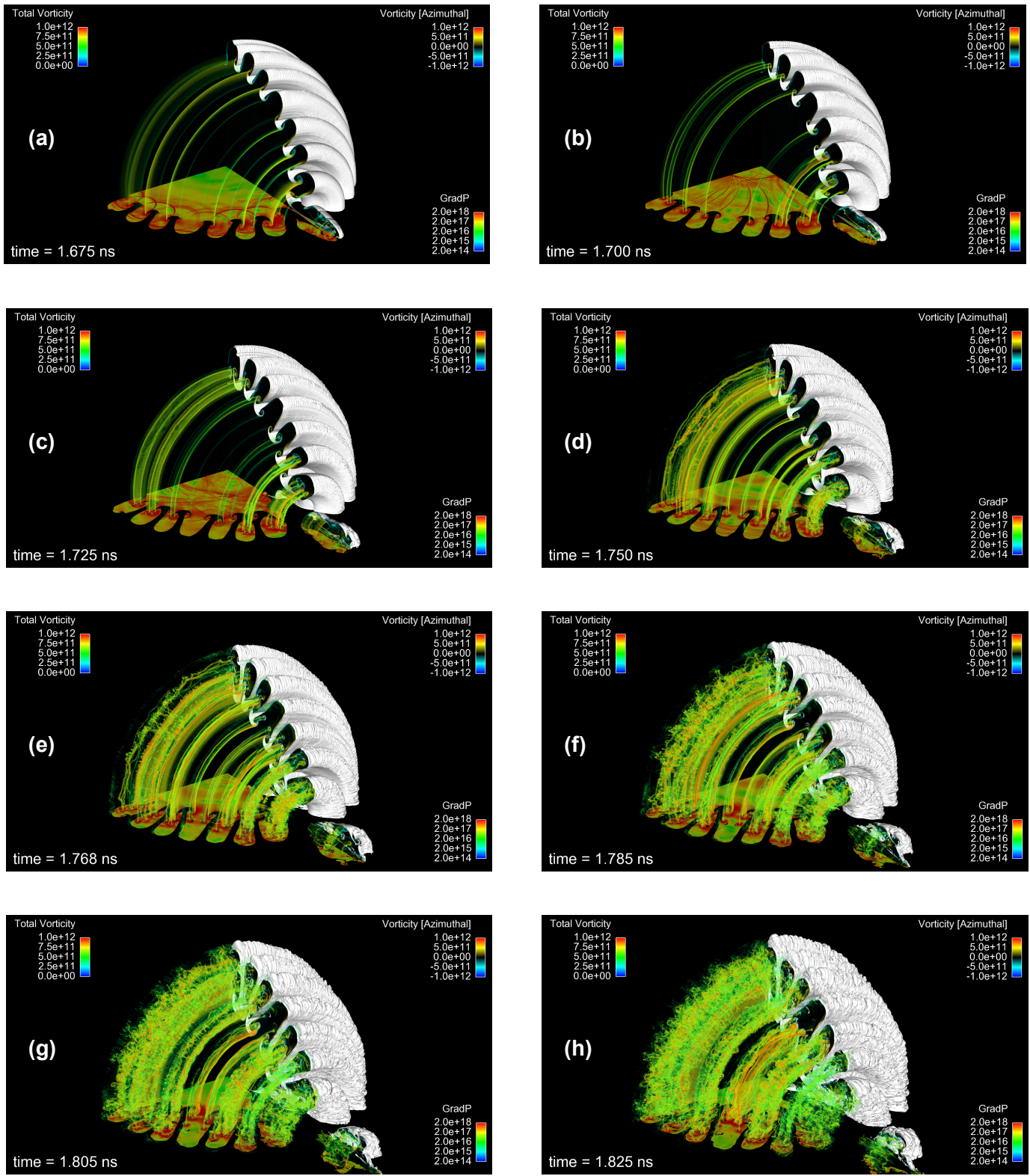

**Fig. 18. Eight time snapshots from the 0.05 μm 3D RAGE simulation of the P30 Omega capsule with a 5% imposed asymmetry $a_{30}$ = 0.05. Full view of the capsule showing the turbulent development of the bubbles.**

Figs. 18(f)–(h) show the appearance and growth of numerous additional long and short wavelength azimuthal instabilities on the vortex rings in the other bubbles, their turbulent evolution, and the development of significant perturbations in the CH/gas interface as the resulting turbulent jets impact the interface.

Fig. 19 illustrates the development of the turbulent jets observed in the 5% case. Fig. 19(a) shows a 2D cut on the lower symmetry plane in the simulation at the time $t = 1.768$ ns slightly after stagnation corresponding to Fig. 18(e). Four panels are shown with the gas colored by the azimuthal vorticity on the top left, by velocity on the top right, by pressure on the bottom left, and by the density on the bottom right. At this time the fingers of shell material converging radially inward have penetrated deeply into the gas constricting the bubble openings. The 6 gigabar pressure peak at the center of the capsule forces a high speed flow of gas radially outward through the constrictions and into the bubbles at speeds of up to 500 km/sec. This flow carries with it the embedded vorticity generated as a result of the gas shock collisions with the incoming perturbed CH/gas interface. As Fig. 19(b) shows, this embedded vorticity is becoming turbulent on a very short timescale as it is carried across the bubbles by the bulk gas flow until it impacts the CH/gas interface as a turbulent jet producing 3D interpenetration of gas and CH. At the time pictured the turbulent jet in the bubble on the far right in Fig. 19(b), the bubble nearest the polar axis, can be seen just beginning to impact the CH/gas interface at the end of the bubble.

This phenomenon is then clearly seen to be a convergence effect since higher convergence leads to more significant penetration of the fingers for a given asymmetry, greater constriction in the bubble openings, and higher relative pressures. Convergence also forces the strong, counter-rotating vortex rings trapped inside the bubbles closer together, further increasing the growth rate of the azimuthal instabilities that lead to the turbulent evolution of the rings. All of these effects should work together to create stronger, and presumably more unstable jets as convergence is increased. It should be noted that the situation in an ICF implosion is somewhat more complicated than that which occurs for a simple gas jet because many regions near the evolving fingers of shell material have significant embedded vortical flow from previous encounters with the reflected shocks. This vortical flow induced earlier in the implosion may become unstable earlier in time, but in any event may be expected to be carried into the bubbles by the jet flow and further contribute to the turbulence in the bubbles.

Comparing the case with a 50% imposed asymmetry to that of the 5% case leads to several observations. First, more of the cavity is hydrodynamically disrupted for the 50% case as would be expected for the larger imposed asymmetry. Second, the 50% case is unstable in 3D earlier in time than the case with 5% imposed asymmetry. In the 50% case we observed turbulent jet formation in the bubbles nearest the polar axis of the capsule shortly after the $2^{nd}$ collision of the reflected gas shock with the CH/gas interface. In the 5% case turbulent jet formation was delayed until after the $3^{rd}$ collision of the reflected gas shock with the CH/gas interface. Again, this is as might be expected since the case with the larger imposed asymmetry generates more enstrophy, as seen from the 2D simulations, but also exhibits stronger convergence effects as the fingers of shell material penetrate earlier and farther into the fuel. This also leads to earlier formation of the gas jets in the more perturbed case. Nonetheless, the behavior of both the 50% and 5% asymmetry cases show turbulence generation and transport occurring in the same generic fashion. More simulations, including ones with fully 3D asymmetries would help to elucidate just how generic this behavior is. That the processes observed in these 3D simulations are seen at low radial convergences of 8 and drive asymmetries as small as 5% suggests that this mechanism of turbulence generation may be significant for many real high convergence ICF implosions even with quite small drive asymmetries.

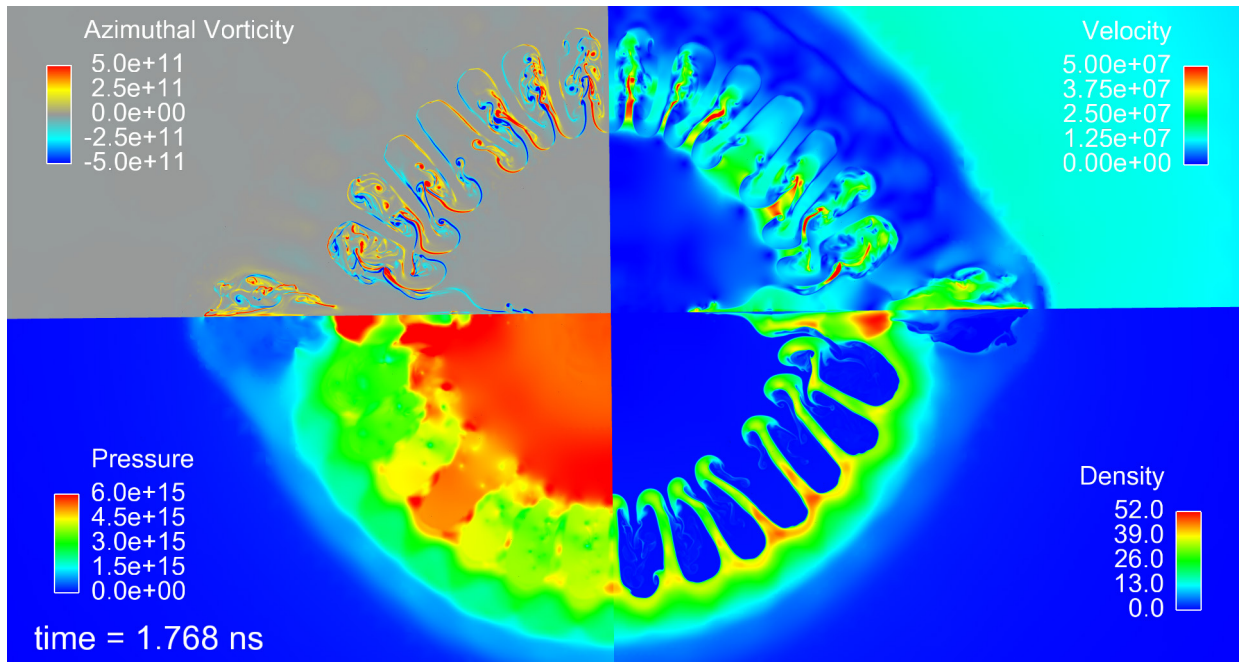

**Fig. 19(a). Four panels showing azimuthal vorticity, velocity, pressure and density on a symmetry plane of the 5% simulation of Fig. 18. Note the 6 gigabar pressure in the capsule center which drives turbulent gas jets outward with velocities of 500 km/sec.**

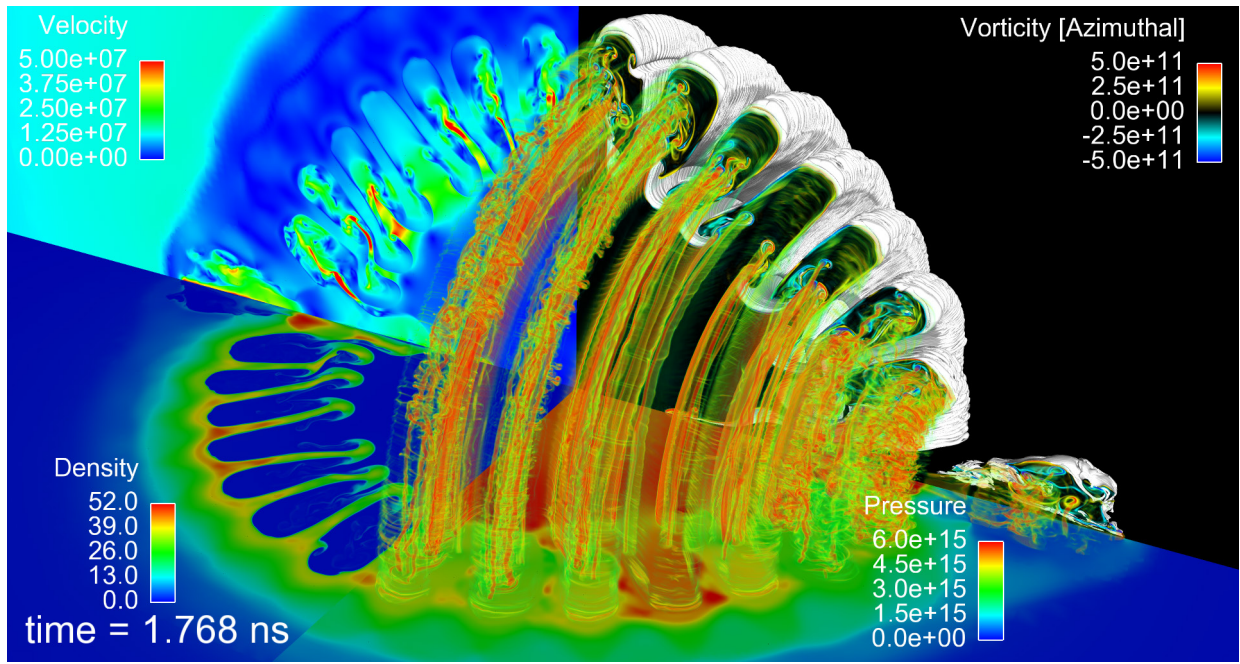

**Fig. 19(b). Another view of the four hydrodynamic quantities of Fig. 19(a) in the 5% 3D RAGE simulation of the P30 capsule with the developing turbulent vorticity shown in a volume-rendered representation. The white surface is, as usual, the CH/gas interface.**

### 2.3 Advantages and Limitations of RAGE for ICF Implosions

Traditionally, ICF implosion simulations have been performed using 2D Lagrangian codes with spatial resolution in the gas so coarse that only the main gas shock is visible in the simulation. Low resolution 2D Lagrangian codes are ill-suited to the problem of asymmetric implosions because the high shear flows which naturally arise in this case are handled poorly if at all by such codes. In contrast, the high resolution adaptive Eulerian technique used in RAGE robustly handles high shear flows and as we have seen our 2D RAGE simulations of asymmetric implosions reveal an enormous wealth of essential detail about the dynamical behavior of shocks and vorticity in the DT gas that remain largely unseen in low resolution 2D Lagrangian calculations.

3D RAGE has an additional important property which makes it well suited for modeling the generation of the gas turbulence in ICF implosions. The numerical hydro scheme being used in RAGE is a monotone integrated large eddy simulation or MILES method[12]. In such methods no explicit subgrid model is required to enforce the physically correct dissipation of energy at small length scales in the modeling of turbulent flows. Rather, the necessary physics of conservation, monotonicity, causality and locality required to correctly model the dissipation are all built into the numerical hydro method itself. For such methods, it has been shown that if sufficient spatial resolution is utilized in 3D to capture most of the energy containing length scales in the flow, then the turbulent structure of the flow is correctly calculated. Hence, we can have some confidence that the structure of the turbulence being observed in our 3D ICF simulations is being correctly calculated.

RAGE also presents some limitations for ICF applications. One of these is the fact that the RAGE simulations being presented here are inviscid. For many ICF implosions the center of the implosion may in fact have a relatively large Spitzer viscosity and a corresponding low Reynolds number because the temperature in this central region is high. As a result vortical structures generated in this region may not lead to significant turbulence generation.

However, near the pusher wall the temperature is likely to be much lower, especially if effects such as electron thermal conduction between the relatively cool pusher and the adjacent DT gas are important. In these regions the viscosity is likely to be much lower and the Reynolds number of the flow much higher than in the center of the DT fuel. As we have seen in our RAGE simulations the majority of the enstrophy generation in our example ICF implosion takes place in these regions adjacent to the CH/gas interface. It is thus reasonable to hope that our inviscid simulation results may accurately reflect how turbulent mixing is really generated in an asymmetrically driven ICF implosion.

It is worth pointing out that some direct experimental evidence exists that changing the viscosity of the DT fuel can have a measureable impact on the performance of the capsule. In experiments on the OMEGA laser, Wilson et. al.[13] reported that adding very small admixtures high Z components to the DT gas had a deleterious effect on the observed performance of the capsule. One effect of these high Z contaminants is to decrease the viscosity of the DT gas and increase its effective Reynolds number. Does the degraded yield observed when high Z contaminants are added to the fuel reflect a decreased viscosity for the gas and an increase in turbulent mixing? This may be an interesting area for future investigations.

An important simplification introduced in the RAGE simulations presented here is that we have considered only the case of an imposed drive asymmetry that is axisymmetric. Our simulations have shown even in this simple case of an axisymmetric perturbation in the pressure drive, that the resultant turbulent flow has a fully three-dimensional character. While it is almost certainly true that drive asymmetries in real ICF implosions are fully three-dimensional in nature, a fully 3D asymmetry in the pressure drive is unlikely to lead to a significantly altered physical picture for how the turbulence arises in the implosion. A fully 3D asymmetry in the pressure drive will still cause non-radial flow in the drive shell leading to density enhancements that will be further amplified by Bell-Plesset related convergence effects. These enhancements will act as seeds for the growth of fully 3D Richtmyer-Meshkov fingers of shell material which will penetrate the gas. At the same time coherent vortical structures will again be created in the gas by the same 3D asymmetries in the pressure drive. The resultant vortex rings will have a more fully three-dimensional character than the highly axisymmetric rings seen in the case of an axisymmetric perturbation but will undergo the same unstable interactions in 3D nevertheless. Hence, the physical picture of an asymmetric drive leading to coherent vortical structures in 3D which in turn lead to turbulence remains relatively unchanged. And this process should be viewed as essentially deterministic since a well-defined final state is achieved with a given pressure drive.

A significant limitation in the use of RAGE for ICF applications is its inability to simulate the implosion of very high convergence systems without the appearance of significant numerical artifacts. In the above 2D RAGE simulations we limited ourselves to implosions of the example OMEGA capsule in which the convergence ratio of the initial to final pusher/gas interface radius was about 8. For this case, as we have shown above, the implosion of a round capsule with a spherically symmetric pressure drive gives a final configuration just before stagnation with only minor numerical artifacts. However, we have noted empirically that for convergence ratios somewhat greater than 11, RAGE gives increasingly noticeable numerical artifacts for a round capsule that is symmetrically driven. Unfortunately, this means that very high convergence problems of great interest like the NIF ignition target with a convergence ratio of the order of 30 cannot be directly simulated using RAGE. However, despite this limitation as we have shown above, the 2D and 3D RAGE simulations provide a great deal of insight into the physical processes that lead to turbulent mix in ICF targets.

We also note more generally that the entire problem of simulating high convergence ICF implosions is unresolved. We do not know under what conditions any given type of simulation can correctly calculate the behavior of asymmetrically driven ICF implosions, especially at high convergence.

## 3. Implications for NIF Ignition

As we have already mentioned above, a direct RAGE simulation of NIF ignition implosions is not possible because the convergence ratio for the NIF ignition target is of order 30 and this is much too high for credible RAGE calculations of the implosion from $t = 0$. Nevertheless, the physical picture of the relationship between drive asymmetry and turbulence generation in an ICF implosion which emerges from the RAGE simulations presented above has potentially important implications for NIF ignition.

The NIF ignition campaign is, of course, an indirect drive experimental program in which a round capsule is enclosed in a cylindrical gold-lined hohlraum designed to absorb and re-radiate the incoming laser energy in order to produce a symmetric radiation drive on the capsule. A recent point design for the capsule itself consists of a germanium-doped CH ablator shell enclosing a cryogenic DT-ice layer with a central region containing DT gas. When the ablator absorbs the drive energy and blows off, the capsule is compressed to very high convergence. Most of the DT fuel remains cold and is compressed to high density while the central region of DT gas, the so-called hotspot, heats to thermonuclear temperatures. When the hotspot ignites, the thermonuclear burn propagates out into the surrounding cold compressed fuel producing a high gain yield.

The details of how this compression is achieved are of some interest. The 20 ns long laser drive pulse has four distinct power peaks that launch four separate drive shocks into the capsule which are designed to coalesce into a single large shock several μm's inside the inner surface of the DT ice layer. This shock then bounces several times between the capsule center and the DT ice layer in order to build up sufficient temperature and $\rho R$ in the hotspot region to achieve ignition.

Our RAGE simulation results raise several physics issues about the basic NIF ignition scheme outlined above. In this discussion the central DT gas region of the NIF hotspot plays the role of the DT gas fill in our example RAGE implosion while the DT ice layer plays the role of the cold pusher shell. We have already seen in our 2D RAGE simulations that at a convergence ratio of 8, even a 1% amplitude for the drive asymmetry was sufficient to cause significant penetration of macroscopic Richtmyer-Meshkov fingers of pusher material into the DT gas whose structure is determined directly by the asymmetry. In the context of the NIF ignition target with its convergence ratio of 30, this suggests that even a few percent amplitude of low to intermediate $\ell$ mode perturbations in the drive symmetry may be sufficient to produce macroscopic fingers of DT ice which penetrate the hotspot. And unlike our simplified RAGE implosion simulations using an axisymmetric drive perturbation, the density enhancements in the DT ice which might arise for the case of an asymmetric drive in NIF could have a more intrinsically 3D character. The convergence effects may be more pronounced for fully 3D drive perturbations because there are two directions of convergence for 3D perturbations versus a single direction of convergence for purely 2D axisymmetric perturbations. Thus, Bell-Plesset effects with 3D asymmetries may lead to longer spikes in the gas earlier in time.

In any case such fingers of cold material in the hotspot would be deleterious for several reasons. First, the effective $\rho R$ of the hotspot would be reduced because the cold fingers absorb alpha particle energy instead of the hotspot, impeding hotspot ignition. Second, the cold fingers displace the hotspot gas to larger radius resulting in lower density and lower temperature, again impeding hotspot ignition. Finally, we have seen in our 3D RAGE simulations that the coherent vortical structures deposited in the bubbles between fingers by collisions of the reflected gas shock with the interface lead to turbulence in 3D on a timescale which is short compared to the implosion time. Hence, we might expect that turbulent mixing of the hotspot gas in the bubbles with the fingers of cold DT ice material could provide a significant impediment to hotspot ignition. Each time the reflected gas shock hits the ice/hotspot interface this effect would be exacerbated. For the NIF ignition target several such collisions occur in the course of the target compression to ignition.

The expected experimental signature of this turbulent mixing phenomenon would be similar to the behavior observed in the high Z experiments of Wilson et al.[13] and the intentionally asymmetric implosion experiments of Rygg et al.[1] where significantly reduced thermonuclear burning was noticed after the reflected gas shock hits the pusher (i.e. the DT ice layer) for the first time. Further degradation will occur with subsequent shock collisions. This possibility is in contrast to the expectations of Wilson et al.[14], where an asymmetric hotspot is expected to show significant deviation in burn rate from the ideal 1D situation only late in time.

One additional aspect of the NIF ignition target is the four drive shocks that are designed to coalesce into one as they exit the ice layer. If the asymmetry of the radiation drive is changing over the time period of the laser pulse due to laser plasma interactions or other effects, then the individual shocks may be asymmetric and may, in fact, have different asymmetries. This is a new more complex possibility than the case of the simple drive pulse with constant drive asymmetry that we considered in our RAGE simulations. In this case, it is possible for additional vorticity production to occur in the gas near the DT ice layer as a result of the interactions of these shocks which could lead to further turbulent mixing of the hotspot and the DT ice layer.

It has been commonly supposed that mixing in the NIF ignition target is dominated by large $\ell$ mode Rayleigh-Taylor instabilities. Our 2D and 3D RAGE simulations suggest, however, an additional possibility. For very high convergence targets like the NIF ignition target, even small asymmetries in the drive can lead to penetration of the hotspot by macroscopic fingers of cold fuel. The same drive asymmetry that leads to these fingers also leads to the coherent vortical structures in the gas whose 3D instability generates turbulent mixing of the hotspot gas with the cold fuel. Experimentally, these two different routes to degraded capsule performance, one due to amplification of small scale perturbations by acceleration-induced instabilities, the other due to turbulent evolution of coherent vortical structures in 3D created by large scale drive asymmetries, may not be distinguishable without spatially and temporally resolved measurements of the hotspot and cold ice layer. However, the overall X ray emission from the assembly may allow inferences about the symmetry of the inside of the implosion. Here the question to ask is whether the overall symmetry is spherical, in which case one might expect the degradation of the burn to result from small scale imperfections in the target. However, for the case in which significant non-spherical features are seen in the X ray emission images even at large radii, it seems fair to infer the presence of significant asymmetry in the actual implosion of the fuel. In this case the yield degradation might be dominated by effects such as the ones seen in our 3D RAGE simulations.

## 4. Conclusions

In this paper we have used high resolution 2D and 3D RAGE simulations to examine the effect of imposing a simple P30 drive asymmetry on the implosion of an example ICF capsule similar in design to plastic capsules fielded in actual OMEGA experiments. In the 2D simulations we showed for the case with zero imposed drive asymmetry that 2D RAGE gave a spherically symmetric implosion with only minor numerical artifacts out to the time of stagnation for a convergence ratio of around 8. For this symmetrically driven case Rayleigh-Taylor growth seeded by numerical perturbations associated with the AMR grid was observed at stagnation as would be physically expected in this case. However, for the cases with a non-zero amplitude for the P30 drive asymmetry the non-radial mass flow induced in the shell by the asymmetric drive leads to a well-defined pattern of density and pressure enhancements in the shell that are further amplified by Bell-Plesset related convergence effects. When the outgoing reflected gas shock collides with the incoming shell, these density enhancements act as seeds for the growth of Richtmyer-Meshkov fingers of shell material which penetrate the DT fuel. In our 2D simulation study varying the amplitude of the asymmetry we showed even for a 1% amplitude for the P30 asymmetry and a convergence ratio of 8 that a clear pattern of 15 fingers of shell material were observed to dominate the structure of the CH/gas interface at stagnation as a result of the action of the P30 asymmetry. At the same time the asymmetry of the drive leads directly to the formation of a pattern of counter-rotating vortex rings in the body of the gas, and the multiple collisions of the reflected gas shock with the perturbed shell result in the formation of strong counter-rotating vortical structures that are trapped in the DT gas bubbles between the converging Richtmyer-Meshkov fingers.

In our linked 3D RAGE simulations with the 50% asymmetry we saw that the counter-rotating vortical structures trapped in the bubbles are unstable in 3D to both short and long wavelength azimuthal instabilities that quickly lead to turbulence in the gas bubbles and complex 3D interpenetration of gas and shell material driven by the turbulent vorticity. We showed that the turbulence observed in our 3D RAGE simulations has all of the properties one expects of fully 3D turbulence including a power spectrum for the kinetic energy with an inertial subrange that is well fit by the Kolmogorov $k^{-5/3}$ law. Finally, we showed that as the spatial resolution of the simulation is increased in 3D, the development of the turbulence occurs more rapidly in time suggesting for sufficiently well resolved 3D simulations that all of the gas bubbles may become fully turbulent before stagnation.

We also demonstrated that even an implosion with 5% imposed asymmetry and a relatively low convergence ratio of 8 displayed similar phenomena near the time of hydrodynamic stagnation suggesting that the hydrodynamics seen in these simulations is quite relevant to real ICF implosions. Furthermore, the unifying concept of turbulent jets was introduced as a way to understand the hydrodynamics near stagnation. This process with the turbulent jets has also been demonstrated to be a convergence effect, possibly providing a qualitative explanation for the experimental observation that high convergence ICF implosions are much more degraded than predicted by current calculations. The NIC failure is a good example of this problem.

It is important to emphasize that the existence of these jets and their turbulent evolution is a convergence effect, making the study of them problematic in planar geometry. Further, this process is not a surface instability but rather an azimuthal instability of the coherent vortical structures in the body of the gas. The surface is surely affected, however, by the impact of the turbulent gas jet upon the pusher surface.

Our convergence scaling calculations in 2D demonstrated the profound effect convergence has on the integrity of our simplified ICF implosion. For instance at the 5% asymmetry level the capsule cannot withstand a convergence much large than 10 without complete hydrodynamic disruption. Also, our results have shown that simple scaling for asymmetry growth due to convergence does not hold once the reflected shocks hit the interface. Therefore conclusions for high convergence ICF implosions must depend largely upon calculations. This presents unique challenges since the real ICF capsule asymmetries are likely fully 3D and time dependent and the calculations themselves have not been demonstrated to give physical answers at high convergence. In the ICF community one often hears the assertion that mix in ICF implosions arises from the amplification of a spectrum of small initial surface imperfections in the capsule by the action of acceleration-induced instabilities like the Rayleigh-Taylor and Richtmyer-Meshkov instabilities. The 3D simulations presented here provide evidence for an additional mechanism for mix in ICF capsule implosions. Asymmetries in the pressure drive on the shell create both fingers of shell material which intrude into the DT fuel as well as the coherent vortical structures in the gas bubbles between the fingers which are unstable in 3D to azimuthal instabilities of the Widnall and Crow type. These azimuthal instabilities quickly lead to the development of turbulence in the DT gas in the form of outward going turbulent jets. The subsequent interaction between this gas turbulence and the CH/gas interface creates a complex 3D structure for this interface, an interpenetration of the gas and metal which represents a type of turbulent mixing in 3D. Thus, in this mechanism the drive asymmetry is responsible for the creation of the coherent vortical structures in the gas and it is the unstable evolution of these structures in 3D which is the source of the gas turbulence. While the detailed turbulent evolution of the vortex rings in 3D is a stochastic process, the nature and distribution of the fingers of shell material and of the coherent vortical structures in the gas are largely deterministic, a result of the detailed asymmetry of the pressure drive.

The 2D and 3D simulation results presented here emphasize the critical role of drive symmetry in ICF ignition attempts like the ongoing ignition experiments at the National Ignition Facility. We have already seen in these simulations that asymmetries in the capsule's pressure drive lead to both turbulence and fingers of shell material in the DT fuel. In the above P30 example we imposed a relatively large drive asymmetry of 50% in an implosion with a relatively low convergence ratio of only around 8 in order to exaggerate these effects for purposes of illustration. But in high convergence systems we might expect to see similar effects for much smaller initial asymmetries in the laser radiation drive. In the NIF ignition target the convergence ratio is around 30, more than three times that of the example OMEGA implosion considered here. In this circumstance it might be expected that only a few percent asymmetry in the laser radiation drive could lead to significant turbulence in the fuel and fingers of DT ice mixing into the central hotspot. An additional complication in the NIF ignition target is the fact that multiple shocks are used to implode the capsule. Any differences in symmetry between these shocks could lead to production of additional vorticity and associated turbulence in the hotspot. Evidence for these phenomena may be seen in the ongoing NIF ignition experiments.

Our RAGE simulations serve to clarify the essential role of convergence and shear in an asymmetrically driven ICF implosion. Pressure drive asymmetries on the capsule lead to non-radial flow in the shell which in turn leads to macroscopic density enhancements in the shell that grow in time as a result of Bell-Plesset related convergence effects. In our 2D RAGE simulations we saw that this Bell-Plesset growth was not just a surface effect but was rather associated with density enhancements throughout the body of the shell. This is in contrast to the typical treatment of the Bell-Plesset effect for incompressible perturbations which allows for uniform compression of the shell only and does not treat the compressible case with strong density gradients in the body of the shell. In the compressible case considered here the Bell-Plesset growth of the density perturbations leads to both radially inward and outward extension of the shell. These macroscopic perturbations act as seeds for the development of Richtmyer-Meshkov fingers of shell material in the gas as a result of multiple interactions with the reflected gas shock. This same interaction results in strong shear flow near the fingers in the form of counter-rotating vortical structures in the gas. Radial convergence pushes both the fingers and their associated vortical flow closer together as the implosion progresses. This radial convergence pushing the fingers together creates nozzles by which the high pressure in the center of the gas creates high speed flows into the gas bubbles trapped between the fingers. This radial convergence also guarantees that the counter-rotating vortical structures will suffer azimuthal instability growth in three dimensions resulting in rapid development of turbulence in the bubbles trapped between the fingers. It is this turbulent growth which leads to interpenetration of shell material and gas that represents fully three-dimensional mixing. The physical picture described here suggests why it is particularly difficult to achieve the ideal 1D yield in a high convergence ICF implosion.

The issue being addressed here is one of fundamental importance: Is the yield degradation observed in ICF implosions typically determined by small scale surface features or by large scale drive asymmetries? We suggest that the large scale coherent structures associated with asymmetries in the pressure drive may play an important role in determining the yield performance of ICF implosion systems, especially at high convergence.

## Acknowledgments

The authors would like to thank the National Nuclear Security Administration's Advanced Simulation and Computing program for its support of this work and for providing computing resources on the ASC Purple and Cielo supercomputers. Thanks also to the members of the RAGE development team at LANL for their help with the use of the RAGE code. Finally, we would like to add our sincere thanks to Anders Grimsrud and the EnSight development team at Computational Engineering International, Inc. for their help with the use of CEI's EnSight visualization software which was used to create all of the visualizations that appear in this paper. This work was performed under the auspices of the U. S. Department of Energy by Los Alamos National Security (LANS), LLC under Contract No. DE-AC52-06NA25396.